\documentclass[journal=jacsat,manuscript=article]{achemso}

\usepackage{chemformula} % Formula subscripts using \ch{}
\usepackage[T1]{fontenc} % Use modern font encodings
\usepackage{mciteplus}
\usepackage{amsmath}
\usepackage{multirow}
\usepackage{multicol}
\usepackage{makecell}
\usepackage{rotating} 
\usepackage{multicol}
\usepackage{xcolor}
\usepackage{booktabs}
\usepackage{graphicx}
\usepackage{amssymb}
\usepackage{hyperref}
\usepackage{caption}
\usepackage{adjustbox}
\usepackage{soul}
\usepackage{gensymb}
\usepackage[flushleft]{threeparttable}
\usepackage{booktabs}
\usepackage{tablefootnote}
\DeclareRobustCommand{\mhl}[1]{%
	\ifmmode\text{\hl{$#1$}}\else\hl{#1}\fi
}
\definecolor{orange}{rgb}{0.82, 0.41, 0.12}
\definecolor{cerulean}{rgb}{0.0, 0.48, 0.65}
\definecolor{ballblue}{rgb}{0.13, 0.67, 0.8}
\definecolor{dodgerblue}{rgb}{0.12, 0.56, 1.0}
\definecolor{oceanboatblue}{rgb}{0.0, 0.47, 0.75}
\author{Camilla Di Mino}
\affiliation[University College London]
{Department of Physics and Astronomy, University College London, Gower Street, London WC1E 6BT, UK}
\author{Adam J. Clancy}
\affiliation[University College London]
{Department of Chemistry, University College London, 20 Gordon Street, London WC1H 0AJ, UK}
\author{Andrea Sella}
\affiliation[University College London]
{Department of Chemistry, University College London, 20 Gordon Street, London WC1H 0AJ, UK}
\author{Christopher A. Howard}
\affiliation[University College London]
{Department of Physics and Astronomy, University College London, Gower Street, London WC1E 6BT, UK}
\author{Thomas F. Headen}
\affiliation[ISIS Neutron Facility]
{ISIS Neutron and Muon Source, Science and Technology Facilities Council, Rutherford Appleton Laboratory, Harwell Campus, Didcot OX11 0QX, UK}
\author{Andrew G. Seel}
\affiliation[ISIS Neutron Facility]
{ISIS Neutron and Muon Source, Science and Technology Facilities Council, Rutherford Appleton Laboratory, Harwell Campus, Didcot OX11 0QX, UK}
\email{andrew.seel@stfc.ac.uk}
\author{Neal T. Skipper}
\affiliation[University College London]
{Department of Physics and Astronomy, University College London, Gower Street, London WC1E 6BT, UK}
\email{n.skipper@ucl.ac.uk}
\phone{+44 (0)207 679 3526}
\fax{+44 (0)207 679 0595}
\title[]{Weak Interactions in Dimethyl Sulfoxide (DMSO) -- Tertiary Amide Solutions: the Versatility of DMSO as a Solvent  
	\footnote{Supporting Information Available}}

\abbreviations{IR,NMR,UV}
\keywords{American Chemical Society, \LaTeX}

\begin{document}

%\begin{tocentry}
%	\includegraphics[width=6.5cm,height=3.5cm]{TOC.png}
%	\label{TOC}
%\end{tocentry}


\begin{abstract}
The structures of equimolar mixtures of the commonly used polar aprotic solvents dimethylformamide (DMF) and dimethylacetamide (DMAc) in dimethylsulfoxide (DMSO) have been investigated via neutron diffraction augmented by extensive hydrogen/deuterium isotopic substitution. Detailed 3--dimensional structural models of these solutions have been derived from the neutron data via Empirical Potential Structure Refinement (EPSR). The inter--molecular Centre-of-Mass (CoM) distributions show that the first coordination shell of the amides comprises $\sim$\,13\,--\,14 neighbours, of which approximately half are DMSO. In spite of this near ideal coordination shell mixing, the changes to the amide-amide structure are found to be relatively subtle when compared to the pure liquids. Analysis of specific intermolecular atom-atom correlations allows quantitative interpretation of the competition between weak interactions in solution. We find a hierarchy of formic and methyl C--H\,$\cdots$\,O hydrogen bonds form the dominant local motifs, with peak positions in the range 2.5\,–\,3.0\,\AA. We also observe a rich variety of steric and dispersion interactions, including those involving the O=C--N amide $\pi$--backbones. This detailed insight into the structural landscape of these important liquids demonstrates the versatility of DMSO as a solvent and the unparalleled resolution of neutron diffraction, which is critical for understanding weak intermolecular interactions at the nanoscale and thereby tailoring solvent properties to specific applications.

\end{abstract}

\section{Introduction}
Polar aprotic liquids are widely used as solvents on both a laboratory and industrial scale, with important applications across a wide range of chemistry, biochemistry and nanoscience.\cite{clancy2015one,chen2012li,yu2020renaissance} In this context, their relevant physicochemical properties include high dipole moments, high relative permittivities, and high boiling points, along with broad electrochemical stability windows when compared to their protic analogues. \cite{1961} This combination of attributes makes these liquids highly effective for solvation of a wide spectrum of ions, small molecules, polymers, and nanostructures \cite{heravi2018beyond,pastoriza1999formation,le2017n,ding2012n,clancy2015one,clancy2018charged}. For example, in electrochemistry their inertness and ability to solvate both metal ions and polymeric co-electrolytes under highly reducing conditions is critical for battery function and stability \cite{yu2020renaissance,chen2012li,chen2017improving}. In addition, polar aprotic liquids provide a unique arena in which to study and tune the fundamental nature of weak inter-molecular interactions, including C--H\,$\cdots$\,O and C--H\,$\cdots\,\pi$ hydrogen bonds and both cyclic and acyclic $\pi$\,--\,$\pi$ effects.

\begin{figure}[h!]
	\includegraphics[width=\textwidth]{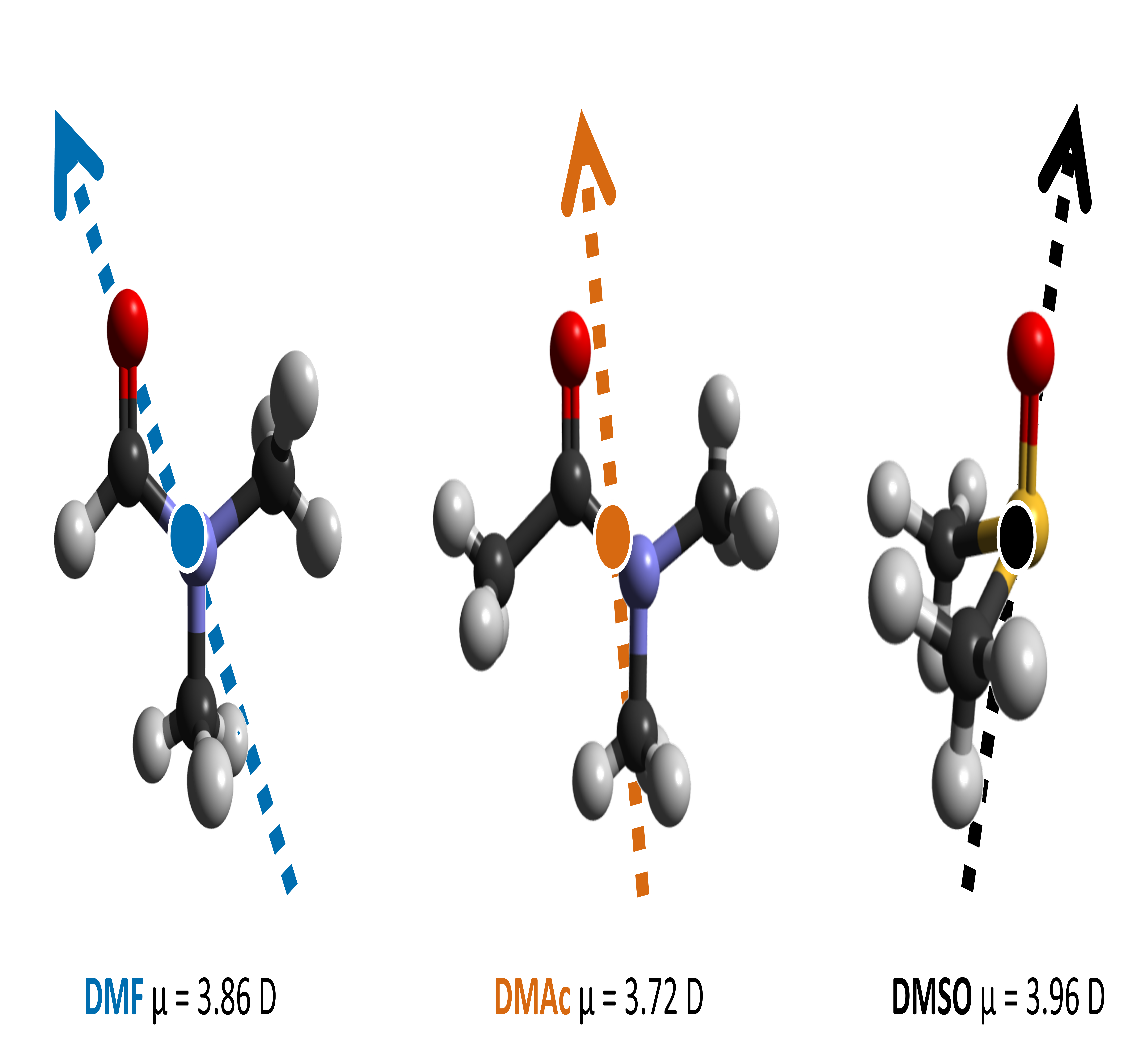}
	\caption{DMF, DMAc and DMSO molecular models. The arrow indicates the direction of the molecular dipole, while the circle through which it passes highlights the centre of mass (CoM) for each species.}
	\label{fgr:dipoles}
\end{figure}

Dimethylformamide (DMF, $\mathrm{Me_{2}NC(=O)H}$) and dimethylacetamide (DMAc, $\mathrm{Me_{2}NC(=O)Me}$) are the simplest aprotic amides, in which the proton donor (protic) N--H groups present in formamide (FA, $\mathrm{H_{2}NC(=O)H}$), N-methylformamide (NMF, $\mathrm{MeHNC(=O)H}$), and N-methylacetamide (NMAc, $\mathrm{MeHNC(=O)Me}$) are replaced by N--Me (Figure\,\ref{fgr:dipoles}). The aprotic nature and the high dipolar character of these amides make them the ideal candidate for studying weak competitive interactions in the liquid state. Both DMF and DMAc are planar acyclic amides, where partial double bond character in the N--C=O framework arises from $\pi$ electron delocalisation that enforces the planarity of the molecule. \cite{mujika2006resonance} DMF and DMAc have similar dipole moments ($\mu$ = 3.86\,D and 3.72\,D respectively) and relative permittivities ($\epsilon_{r}$ = 36.8  and 37.8 at 20\,\degree C respectively, Table\,\ref{tb:TDparameters}) which lead to strong dipole--dipole interactions and relative orientational effects in the liquid structure. \cite{basma2019liquid}
Both molecules are regarded as weak Lewis bases, with donor numbers (DN) of 26.6 $\mathrm{kcal\, mol^{-1}}$ and 27.8 $\mathrm{kcal\, mol^{-1}}$ and acceptor numbers (AN) of 16.0 $\mathrm{kcal\, mol^{-1}}$ and 13.6 $\mathrm{kcal\, mol^{-1}}$ respectively for DMF and DMAc, Table\,\ref{tb:TDparameters}.\cite{laurence2009lewis} Furthermore, the presence of a C(=O)--H group in DMF raises the possibility of hydrogen bonding by a weakly donating formic H atom. On a practical level, this functionality also means that while DMF is one of the most heavily used solvents for chemical synthesis, it can react under highly basic conditions and with strong reducing and chlorinating agents. DMAc is usually more inert, and so has complementary applications for example in the production of pharmaceuticals and polymers.\cite{mujika2006resonance} 

Neutron diffraction studies of liquid DMF and DMAc have shown well-defined local structures.\cite{basma2019liquid} For both cases, the coordination number of molecules in the first solvation shell is found to be around 13, with a clear second shell also present. In DMF, weak C(=O)--H\,$\cdots$\,O hydrogen bonds are observed that are thought to be electrostatic in nature. In DMF the first solvation shell shows the expected preference for anti-parallel dipole orientation between molecules, while in DMAc parallel dipoles maximise dispersion forces between the $\pi$-delocalised O=C--N backbones and methyl groups.\cite{basma2019liquid}
These results are consistent with Raman, IR spectroscopy and molecular dynamics studies of liquid DMF, all of which reported a highly structured first solvation shell with weak hydrogen bonding from carbonyl and methyl H atoms, in line with the 'relatively short' C–H\,$\cdots$\,O=C contacts seen earlier by gas-phase electron diffraction. \cite{schultz1993molecular,schoester1995comparison,cordeiro1999study,macchiagodena2016accurate,park2009intermolecular,lei2003structures}

Dimethylsulfoxide (DMSO, $\mathrm{Me_{2}S=O}$) is a pyramidal molecule with high dipole moment ($\mu$ = 3.96\,D), weak Lewis base character (DN = 29.8 $\mathrm{kcal\, mol^{-1}}$ and AN = 19.3 $\mathrm{kcal\, mol^{-1}}$), and a remarkably high permittivity for an aprotic solvent ($\epsilon_{r}$ = 47.2 at 20\,\degree C). Its unique properties are due to the combination of a soft lone pair on the sulfur atom and the strong polarisation of the S=O bond. DMSO, like DMF and DMAc, is miscible with water and many organic solvents, and has an unique ability to solvate a wide range of chemical species from apolar hydrocarbons to entirely dissociated salts. DMSO is therefore extremely important in processing and technology. \cite{doi:https://doi.org/10.1002/9783527805624.ch7} Moreover, DMSO is able to penetrate human skin with a non-destructive effect on tissues and is a keystone protectant in cryobiology.\cite{awan2020dimethyl} Additional theoretical interest stems from the long-standing question of how to best represent the sulfoxide bond: $\mathrm{S^{+}}$--\,$\mathrm{O^{-}}$ rather than a formal S=O double bond? \cite{moffitt1950nature,cruickshank19611077,clark2008dimethyl} Structural studies of liquid DMSO by X--ray and neutron diffraction have reported nearest neighbour coordination numbers in the range 11.5\,–\,13.8 and have provided evidence for short-range antiparallel alignment of dipoles, with head-to-tail ordering at longer distances. In addition, weak methyl hydrogen to oxygen inter-molecular contacts were observed at distances of approximately 3.4\,\AA, but as such these probably do not constitute hydrogen bonds. \cite{luzar1993combined,mclain2007investigations,soper2012computer,megyes2004x}

While the bulk physicochemical properties of DMF, DMAc and DMSO are therefore similar (Table\,\ref{tb:TDparameters}), the pure liquids exhibit contrasting local structures. This latter point immediately raises the question as to which interactions will dominate in mixtures of these molecules, and in particular how, and to what extent, DMSO is accommodated within the local solvation environments of DMF and DMAc (and \textit{vice versa}). 

Previous studies of mixtures the polar protic solvent NMF in DMSO by neutron diffraction point to the formation of a strong N--H\,$\cdots$\,O=S hydrogen bond between the protic amine group of NMF and the oxygen of DMSO at 1.6\,\AA. \cite{cordeiro2011investigation} The latter distance is considerably shorter than the typical strong hydrogen bonding in the liquid state: taking the interaction between water molecules as an example, the first O--H\,$\cdots$\,O contact is found at 1.85\,\AA \space at ambient conditions.\cite{soper1986new,soper1997site} NMF-DMSO hydrogen bonding is more similar in length to liquid HF, one of the shortest hydrogen bonds reported in a liquid.\cite{mclain2004structure} As a consequence, NMF and DMSO molecules form very stable dimers in the mixtures. This in turn results in well-organized first and second solvation shells with a preference for heteromolecular NMF–DMSO hydrogen bonds, rather than homomolecular NMF–NMF. \cite{zhou2016evidences,cordeiro2011investigation,chebaane2012intramolecular,borges2013hydrogen,ludwig1995temperature} NMF, though, is a protic, highly polar solvent ($\mu$ = 3.86\,D), with extremely high relative permittivity ($\epsilon_r$\,=\,181 at 25\,\degree C) and the ability to act as both proton donor and acceptor via its N--H and C=O groups. In clear contrast to NMF, DMF and DMAc are aprotic solvents that only form weak hydrogen bonds via the C--H and methyl groups. As such, they will pose a very different conundrum for DMSO as a co-solvent when they are compared with NMF. 

In this study we have used neutron diffraction in conjunction with isotopic substitution of hydrogen (H) by deuterium (D) to study both the pure liquid amides DMF and DMAc and equimolar 50\,:\,50 mixtures of these with DMSO to understand the role of weak intermolecular interactions, such as weak hydrogen bonding and dispersion forces, on a molecular level and to reveal the role of DMSO as co-solvent in a aprotic environment. The use of the Empirical Potential Structure Refinement (EPSR) computations has allowed us to uncover the 3--dimensional site--site correlations in these systems.\cite{soper1996empirical}

\begin{table}
	\resizebox{\textwidth}{!} {
		\begin{tabular} {lccccccc}
			\toprule
			&$\mu$ /D  &M.P. /\,\degree C & B.P. /\,\degree C& $\rho$ /g\,$\mathrm{cc^{-1}}$ & $\epsilon_{r}$& Molecular Volume /\AA $^{3}$& DN -- AN \cite{laurence2009lewis} / $\mathrm{kcal\, mol^{-1}}$  \\
			\midrule
			\color{cerulean}{\textbf{DMF}} 
			&3.86 &-61 &153&0.944 & 36.7 & 128 & 26.6 -- 16\\
			%	& 3.86 & -61 & 153 & 0.92 &0.944& 128 & ​36.7\\
			\midrule
			\color{orange}{\textbf{DMAc}}
			&3.72 &-20 &165 &0.940 & 37.8 & 154 & 27.8 -- 13.6\\
			\midrule
			\textbf{DMSO} &3.96&19 &189&1.10 &46.7 & 118& 29.8 -- 19.3\\
			\bottomrule
	\end{tabular}}
	\caption{Selected physicochemical properties of the liquids DMF, DMAc and DMSO. }
	\label{tb:TDparameters}
\end{table}
\section{Theoretical Basis}
The function of interest which can be extracted from a neutron diffraction measurement is know as the total structure factor, $F(Q)$, which can be written as:

\begin{equation}
	F(Q)=\sum_{\alpha,\beta \geqslant \alpha }{(2-\delta_{\alpha \beta})\,b_{\alpha} b_{\beta} c_{\alpha} c_{\beta}\, (S_{\alpha\beta}(Q) - 1 )}
	\label{eqn:tsf}
\end{equation} 
where $c_{\alpha} $ and $c_{\beta}$, $b_{\alpha} $ and $b_{\beta} $ are respectively the fractional concentrations of the atomic species $\alpha$ and $\beta $ and the (isotope dependent) coherent neutron scattering lengths, $ Q=4\pi\frac{\sin\theta}{\lambda} $ is the magnitude of the neutron scattering vector, and $S_{\alpha\beta}(Q) $ the Faber-Ziman partial structure factor for any two types of atoms. There is, therefore, a unique \textit{F(Q)} for each isotopic composition (isotopologue) of a sample. In particular, we can exploit the difference in sign and magnitude between the coherent neutron scattering lengths of hydrogen ($b_H =$ --\,3.74\,fm) and deuterium ($b_{D} =$ 6.72\,fm) to distinguish between specific sites in a molecule and thereby measure multiple distinct $F(Q)$s. This approach constrains the overall structure refinement so that we can interrogate the individual site-site correlations needed to describe the structure of a complex liquid. \cite{sears1992neutron,squires1996introduction,terban2021structural} 

The partial structure factors are related to the partial radial distribution functions (RDFs), $g_{\alpha\beta}(r)$, via Fourier transformation:
\begin{equation}
	S_{\alpha\beta}(Q)= 1 + 4\pi \rho \int^{\infty}_{0} dr \; r^{2} \,[g_{\alpha \beta}(r) - 1 ] \,\frac{sin(Qr)}{Qr}.
	\label{eqn:pdf}
\end{equation} 
where $\rho$ is the atomic number density and \textit{r} is the distance between two species $\alpha$ and $\beta$. The $g_{\alpha\beta}(r)$s represent the probability density of finding, by spherical averaging, an atom of species $\beta$ at distance \textit{r} from an atom of species $\alpha$ chosen as origin of the reference system. These functions therefore contain important site-specific structural information of the sample \cite{terban2021structural}. In a liquid system, $g_{\alpha\beta}(r)$ tends to 1 at large values of \textit{r}.

In order to quantify the average coordination number, $N_{\alpha\beta}(r_{0})$, of sites of type $\beta$ in proximity to a site of type $\alpha$ up to a maximum distance $r_{0}$, one can integrate the partial radial distribution function $g_{\alpha \beta}(r)$ over separation distance, \textit{r}:

\begin{equation}
	N_{\alpha\beta}(r_{0}) = 4\pi \int_{0}^{r_{0}} g_{\alpha\beta}(r)\, \rho_\beta \, r^{2} \, dr
	\label{eqn:coord}
\end{equation}
where $\rho_{\beta}$ is the number density of species $\beta$ and $r_{0}$ is the maximum distance of integration. By definition, the first coordination number gives the average number of sites of species $\beta$ present in a sphere of radius $r_{0}$ centred on a site of species $\alpha$. Traditionally, the upper limit of the integral in Equation \ref{eqn:coord} is the position of the first minimum of the partial $g_{\alpha \beta}(r)$. Alternatively, the cumulative coordination numbers can be plotted as a function of the distance $r$ from the central species.  

Beyond a one-dimensional analysis, the Spatial Density Functions (SDFs) are a three-dimensional map of the density of neighbouring molecules around a central molecule as a function of angular distance, \textit{r}, and angular position $\theta$. The SDFs therefore represent regions of space around a central molecule that are most likely to be occupied by a molecule of the same or another species at a given distance \cite{steele1963statistical,hansen1988theory}. 
	
\section{Experimental Details}
Experimental data have been acquired at the Near and InterMediate Range Order Diffractometer (NIMROD) at the ISIS Neutron and Muon Source (Didcot, UK) across a Q range of  0.05\,\AA$^{-1} -$ 50\,\AA $^{-1} $ \cite{bowron2010nimrod}. DMF, DMAc and DMSO and their isotopes were purchased from Sigma Aldrich with purities $\geq 99.5\%$ and handled under inert atmosphere. For pure DMF and DMAc, fully hydrogenated, fully deuterated and a 50:50 mixture of hydrogenated and deuterated liquids were loaded into $\mathrm{Ti_{0.68}\,Zr_{0.32}}$ null scattering cells to give a total of 3 isotopically distinct samples for each liquid amide. To produce 50\,:\,50 amide/DMSO mixtures, the anhydrous liquids were mixed to obtain 7 isotopically distinct samples for DMF/DMSO and 5 for DMAc/DMSO, as summarised in Table\,\ref{samples}. 
The samples were inserted into flat-plate null coherent scattering titanium/zirconium cells, with 1 mm sample and wall thicknesses. Each composition was run for a minimum of 2 hours at 298\,K. Data for the pure liquid DMF and DMAc have been reanalysed using the same methods and protocols as the mixture to allow for a more rigorous comparison. \cite{basma2019liquid} To allow data correction and calibration, scattering data were also collected from the empty instrument, empty sample cells, and an incoherent scattering vanadium-niobium reference slab of thickness 3\,mm. Reduction of the experimental data, including absolute normalisation, background subtraction, and multiple and inelastic scattering corrections, has been conducted using standard procedures as implemented within the Gudrun package. \cite{soper2011gudrunn} 
\begin{table}[h!]
	\begin{threeparttable}
	\begin{tabular} {lcccc}
		&\color{cerulean}{\textbf{Pure DMF}}& \color{orange}{\textbf{Pure DMAc}}& \color{cerulean}{\textbf{DMF}}\color{black}\textbf{\,:\,DMSO 50\,:\,50}& \color{orange}\textbf{DMAc}\,\color{black}\textbf{\,:\,DMSO 50\,:\,50} \\
		1&$\mathrm{D_{7}}$&$\mathrm{D_{9}}$&$\mathrm{D_{7}}$:$\mathrm{D_{6}}$&$\mathrm{D_{9}}$:$\mathrm{D_{6}}$\\	
		2& $\mathrm{H_{7}}$& $\mathrm{H_{9}}$&$\mathrm{D_{7}}$:$\mathrm{H_{6}}$&$\mathrm{D_{9}}$:$\mathrm{H_{6}}$\\
		3&$\mathrm{H_{7}}$/$\mathrm{D_{7}}$&$\mathrm{H_{9}}$/$\mathrm{D_{9}}$&$\mathrm{D_{7}}$:$\mathrm{H_{6}}$/$\mathrm{D_{6}}$&$\mathrm{H_{9}}$/$\mathrm{D_{9}}$:$\mathrm{H_{6}}$/$\mathrm{D_{6}}$\\
		4&-&-&$\mathrm{H_{7}}$/$\mathrm{D_{7}}$:$\mathrm{D_{6}}$&$\mathrm{H_{9}}$:$\mathrm{D_{6}}$\\
		5&-&-&$\mathrm{H_{7}}$/$\mathrm{D_{7}}$:$\mathrm{H_{6}}$/$\mathrm{D_{6}}$&$\mathrm{H_{9}}$:$\mathrm{H_{6}}$\\
		6&-&-&$\mathrm{H_{7}}$:$\mathrm{D_{6}}$&-\\
		7&-&-&$\mathrm{H_{7}}$:$\mathrm{H_{6}}$&-\\		
		
	\end{tabular}
	
	\caption{List of the isotopically distinct samples run in the neutron experiments$^{*}$\,.}
		\label{samples}
	\begin{tablenotes}
		
	\item[$*$]{\color{cerulean}{\textbf{DMF}}\color{black}: $\mathrm{D_{7}= DC(O)N(CD_{3})_{2}}$, $\mathrm{H_{7}= HC(O)N(CH_{3})_{2}}$;

	\item[$*$]\color{orange}{\textbf{DMAc}}\color{black}: $\mathrm{D_{9}= D_{3}C(O)N(CD_{3})_{2}}$, $\mathrm{H_{9}= H_{3}C(O)N(CH_{3})_{2}}$\color{black};

	\item[$*$]\textbf{DMSO}: $\mathrm{D_{6}= (CD_{3})_{2}SO}$, $\mathrm{H_{6}=(CH_{3})_{2}SO}$.}
	
	\end{tablenotes}
\end{threeparttable}

\end{table}

\section{Computational Details}
The EPSR method consists of a classical Monte Carlo molecular simulation which takes initial seed potentials for modelling pairwise interactions, and subsequently refines these through the incorporation of an empirical potential. This empirical potential is calculated with reference to any mismatch between the experimental and simulated data, until a satisfactory agreement between the calculated structure factors and the measured neutron scattering data is reached. In this manner, a three-dimensional structural model of the system can be obtained which is consistent with the experimental data. 
 
\begin{figure}[h!]
	\includegraphics[width=\textwidth]{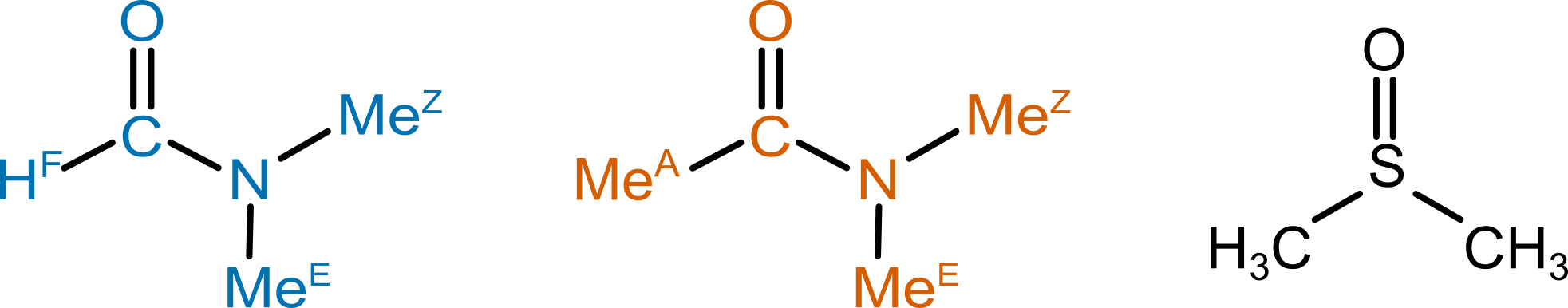}
	\caption{DMF, DMAc and DMSO molecular models with relevant atomic sites labelled. }
	\label{fgr:model3}
\end{figure}

The inter-molecular potential between two atomic sites $\alpha$ and $\beta$ is modelled in EPSR via a Lennard--Jones 12--6 function plus a Coulombic term: 

\begin{equation}
U_{\alpha \beta}(r_{ij}) = 4 \epsilon_{\alpha \beta} \Bigl[\Bigl(\frac{\sigma_{\alpha \beta}}{r_{ij}}\Bigr)^{12}-\Bigl(\frac{\sigma_{\alpha \beta}}{r_{ij}}\Bigr)^{6}\Bigr]+ \frac{q_{\alpha} q_{\beta}}{4 \pi \epsilon_{0} r_{ij}}
	\label{eqn:LJ}
\end{equation}

where $q_{\alpha,\beta}$ are the atomic partial charges, $\epsilon_{\alpha \beta}$ and $\sigma_{\alpha \beta}$ are the well depth parameter and the range parameter respectively and are given by the Lorentz--Berthelot mixing rules in terms of their values of the individual atoms.\cite{lorentz1881ueber} All the molecules were generated in $Avogadro$ and their geometry is optimised for 500 steps using the MMFF94 force field. \cite{halgren1996merck,hanwell2012avogadro} (see Table S1 for intramolecular parameters) The intermolecular seed potentials are taken from the OPLS series of force-fields. \cite{jorgensen1985optimized,jorgensen1996development,zheng1996molecular,yan2017improved}
The cubic EPSR boxes of side-length 44.80\,\AA\space and 47.58\,\AA \space contain 700 molecules for pure liquid DMF and DMAc while cubic boxes of side 49.77\,\AA\space and 51.63\,\AA \space containing 1000 molecules of which 500 are DMF/DMAc and 500 are DMSO. The atomic number densities for the four systems are 0.0934\,atoms/\AA$^{3}$ and 0.0975\,atoms/\AA$^{3}$ for the DMF and DMAc pure liquid and 0.08925\,atoms/\AA$^{3}$ and 0.09080\,atoms/\AA$^{3}$ for DMF/DMSO and DMAc/DMSO mixtures. These values are obtained from a weighted average of the relevant bulk densities of the pure liquids and are verified by reference to the overall scattering levels of the experimental data. The labels assigned to atomic sites of the DMF, DMAc and DMSO molecules are shown for clarity in Figure\,\ref{fgr:model3}.

\begin{table}[h!]
	
	\centering
	\begin{tabular}{lcccc}
		\toprule
		& & $\sigma $ /\AA& $\epsilon $ /kJ\,$\mathrm{mol^{-1}}$ &q /e \\
		\midrule
		\multirow{8}*{\color{cerulean}\rotatebox[origin=c]{90}{\textbf{DMF \cite{jorgensen1985optimized}}}}
		& \color{cerulean} $\textbf{O}$ & 2.96 & 0.87864	& -0.50\\
		& \color{cerulean} $\textbf{C}$ & 3.80	&0.48116&  0.50 \\
		& \color{cerulean} $\textbf{H}$ &- & -&	- \\
		& \color{cerulean} $\textbf{N}$ & 3.25	&0.71128& -0.57\\
		& \color{cerulean} $\mathbf{C^{Z}}$ &3.80	&0.71128& 0.285 \\
		& \color{cerulean} $\mathbf{H^{Z}}$ &-	&-& -  \\
		& \color{cerulean} $\mathbf{C^{E}}$ &3.80	&0.71128& 0.285  \\
		& \color{cerulean} $\mathbf{H^{E}}$ &-	&-& -  \\
		\midrule
		\multirow{9}{*}{\color{orange}\rotatebox[origin=c]{90}{\textbf{DMAc \cite{jorgensen1996development}}}} 
		& \color{orange} $\textbf{O}$ & 2.96 & 0.87864	& -0.50 \\
		& \color{orange} $\textbf{C}$ & 3.75	&0.43932&  0.50  \\
		& \color{orange} $\mathbf{C_{A}}$ &3.50 & 0.25104&	-0.18 \\
		& \color{orange} $\mathbf{H_{A}}$ &2.50 & 0.12552&	0.06 \\
		& \color{orange} $\textbf{N}$ &3.25	&0.71128& -0.14  \\
		& \color{orange} $\mathbf{C^{Z}}$ &3.50	&0.27614& -0.11  \\
		& \color{orange} $\mathbf{H^{Z}}$ &2.50	&0.12552& 0.06  \\
		& \color{orange} $\mathbf{C^{E}}$ &3.50	&0.27614& -0.11  \\
		& \color{orange} $\mathbf{H^{E}}$ &2.50	&0.12552& 0.06  \\
		\midrule
		\multirow{4}*{\rotatebox[origin=c]{90}{\textbf{DMSO \cite{zheng1996molecular}}}} 
		&  $\textbf{O}$ &2.93& 1.1712	& -0.459  \\
		&  $\textbf{S}$ &3.56&1.6522	& 0.139  \\		
		& $\textbf{C}$ & 3.81 & 0.66923 & 0.160 \\
		& $\textbf{H}$ &-& -	& - \\
		
		\bottomrule
	\end{tabular}
	
	\caption{Lennard-Jones parameters and charges for DMF\cite{jorgensen1985optimized}, DMAc \cite{jorgensen1996development} and DMSO \cite{zheng1996molecular,yan2017improved} from top to bottom. Atomic sites correspond to the labelling of Figure \ref{fgr:model3}.} 
	\label{LJparameters}
\end{table}

\section{Results and Discussion}
\begin{figure}[h!]
	\includegraphics[width=\textwidth]{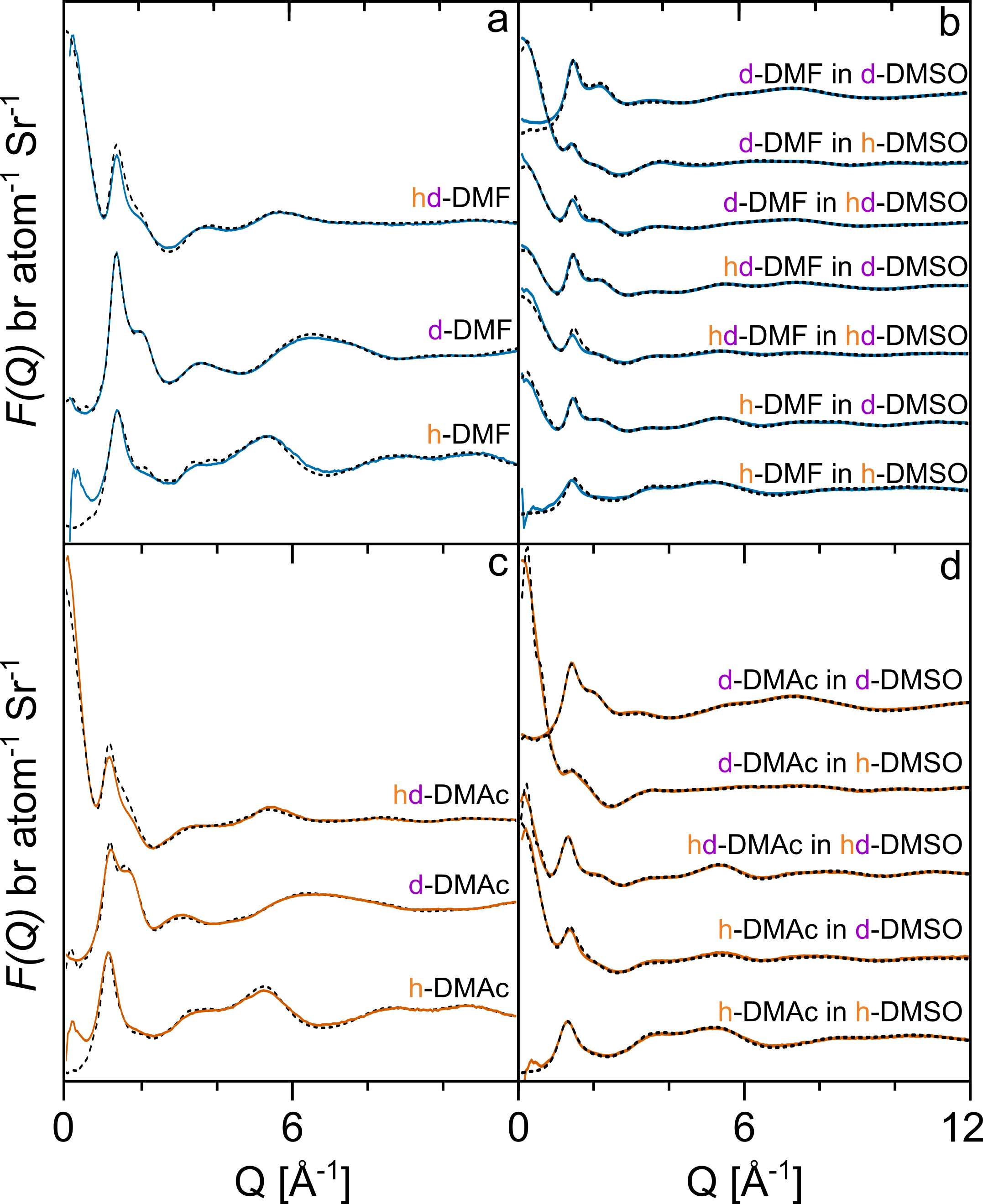}
	\caption{Experimental (solid-line) and modelled (dashed-line) neutron diffraction total structure factors, \textit{F(Q)}, for pure liquid DMF\,(a) and DMF/DMSO\,(b) mixture (left) and pure liquid DMAc\,(c) and DMAc/DMSO\,(d) mixture (right).}
	\label{fgr:fit}
\end{figure}
The isotopically distinct experimental neutron diffraction structure factors, $F(Q)$, are plotted for the pure liquid amides and their DMSO mixtures against the EPSR model in Figure\,\ref{fgr:fit}. Excellent agreement between the experimental data and model has been achieved for each data-set; the small discrepancies at low--$Q$ in fully hydrogenated samples such as h-DMF and h-DMAc are attributed to a residual presence of inelastic and multiple scattering events \cite{soper2009inelasticity,soper2011gudrunn}. The rise in low--$Q$ scattering for samples such as hd-DMF and hd-DMAc can then be attributed to the fact that these liquids are comprised of a mixture of fully hydrogenated and fully deuterated molecules, Table\,\ref{samples}. This isotopic partitioning on individual molecules leads to a genuine rise in elastic scattering that is well captured by the EPSR model. These neutron diffraction data (Figure\,\ref{fgr:fit} solid-line) show clearly that the liquid amides are fully miscible with DMSO at this concentration, as there is an absence of any residual low--$Q$ signal that would indicate the presence of homomolecular clustering. This observation can be confirmed by examining the molecular Centre-of-Mass (CoM) radial distribution functions, $g_{CoM-CoM}(r)$ obtained from the EPSR model.
      
\begin{figure}[h]
	\includegraphics[width=\columnwidth]{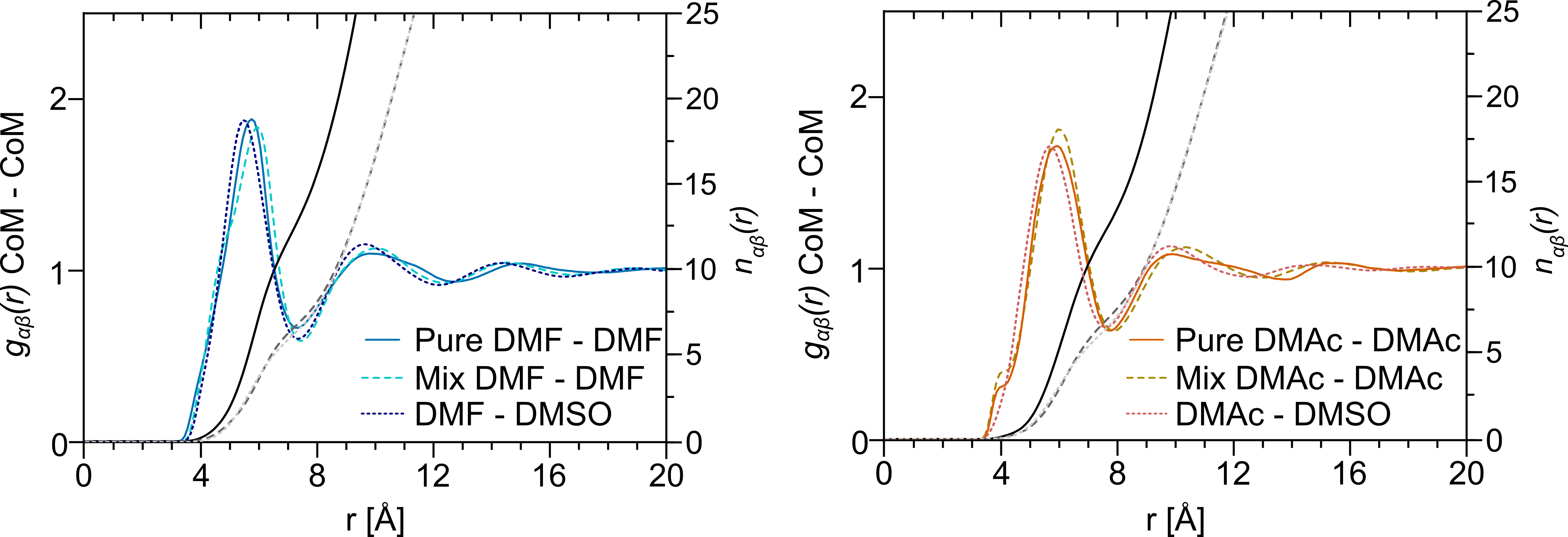}
	\caption{CoM--CoM inter-molecular partial radial distribution functions, $g_{CoM-CoM}(r)$, and cumulative coordination numbers, $N_{CoM-CoM} (r)$, for DMF (left) and DMAc (right). Note that the RDFs for the pure liquids and the mixtures are very similar, showing that DMSO does not disrupt the amide--amide correlations as it infiltrates the local coordination. In addition, we observe clear second and third solvation shells which, if anything, become more ordered in the presence of DMSO.}
	\label{fgr:g(r)CoM}
\end{figure}

\begin{table}[h!]
	\centering
	\resizebox{\textwidth}{!} {
		\begin{threeparttable}
		\begin{tabular}{lcccc}
			\toprule
			& \makecell{$\mathrm{1^{st}}$ Peak /\AA \\ Pure/Mix}& \makecell{$\mathrm{2^{nd}}$ Peak /\AA \\ Pure/Mix} & \makecell{Integration Limit /\AA \\ Pure/Mix} & \makecell{ Coordination Number ($\mathrm{\pm}\,0.1$) \\ Pure/Mix}\\
			\midrule
			
			\color{cerulean}{\textbf{DMF -- DMF}} & 5.76/5.90 & 9.97/10.05 & 7.46$^{*}$/7.46$^{*}$	& 13.4/7.1\\
			\color{orange}{\textbf{DMAc -- DMAc}} & 5.94/6.00 & 9.92/10.27 & 7.87$^{*}$/7.87$^{*}$	& 13.0/7.4\\
			
			\midrule 
			
			\color{cerulean}{\textbf{DMF -- \color{black}{DMSO}}} & 5.50 & 9.64 & 7.36	& 6.6\\
			\color{orange}{\textbf{DMAc -- \color{black}{DMSO}}}
			& 5.68 & 9.84 & 7.62	& 6.5\\
			\textbf{DMSO -- DMSO \cite{megyes2004x}}  & 5.28& - & 7.28	& 13.8 \\
			\bottomrule
	\end{tabular}
		 \begin{tablenotes}
		\item[$*$] The integration limit is set to the first minimum of the mix to allow more direct comparison between the corresponding coordination numbers.
		
	\end{tablenotes}
\end{threeparttable}
} 	
\caption{CoM-CoM first and second peak positions, integration limits and coordination numbers in the pure liquid amides and the 50\,:\,50 mixtures with DMSO.}
\label{tb:coordCoMamide}
\end{table}

\begin{figure}[h]
	\includegraphics[width=\columnwidth]{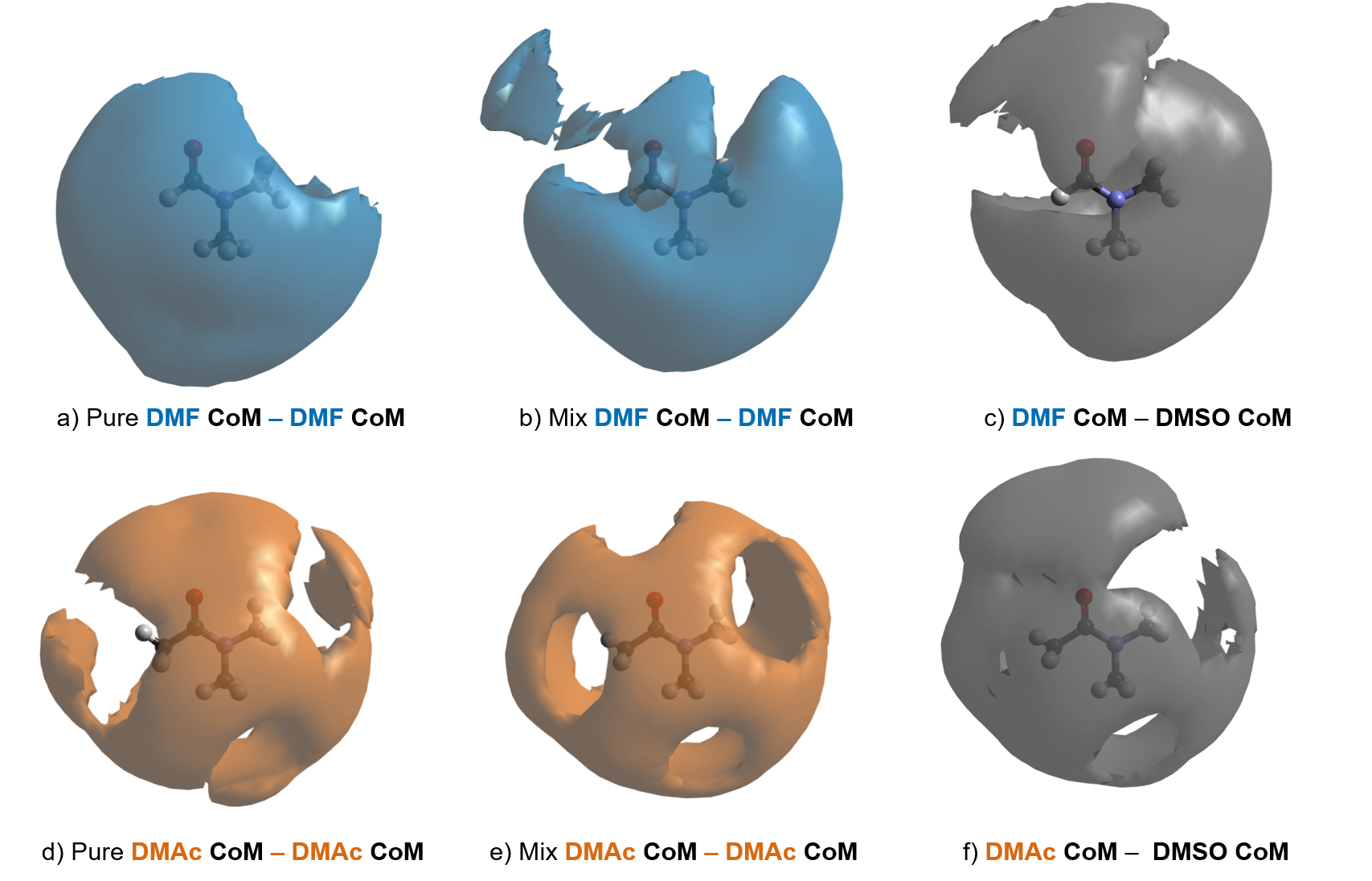}
	\caption{CoM--CoM spatial density functions (SDFs) representing the 25\% most likely configuration of: (a,b) the DMF molecule CoM around another DMF up to 7.5\,\AA\space from the DMF Centre--of--Mass (CoM) in bulk liquid amide (left) and in the mixture with DMSO (right); (d,e) the DMAc molecule CoM around another DMAc up to 7.9\,\AA\space from the DMAc CoM in bulk liquid amide (left) and in the mixture with DMSO (right); (c,f) the DMSO molecule CoM around a DMF and a DMAc up to 7.5\,\AA\space and 7.9\,\AA\space distance from the DMF and DMAc CoM respectively.}
	\label{fgr:CoM-SDFs}
\end{figure}

Figure\,\ref{fgr:g(r)CoM} and Table\,\ref{tb:coordCoMamide} present the radial distribution functions (RDFs) for the CoM amide--amide interactions, $g_{CoM-CoM}(r)$, and the cumulative coordination number, $N_{CoM-CoM} (r)$, as a function of the separation distance \textit{r}. These RDFs generated by EPSR are compared with those obtained by classical Monte Carlo simulation in Supporting Information Section S5. We see immediately from these functions that the local structure of liquid DMF and DMAc, as depicted by the CoM--CoM RDFs, is almost unaltered by the presence of DMSO in the mixtures. Moreover, the plotted and tabulated cumulative coordination numbers confirm that approximately half of the $\sim$\,13\,--\,14 first shell amide molecules in the pure liquids are replaced by DMSO in the mixtures. This corresponds to near ideal coordination shell mixing at the molecular level. In addition, the RDFs provide clear evidence for a second and third solvation shell. 

Our CoM--CoM RDF data do reveal subtle structural differences in peak features and positions, for example a small shortening of the heteromolecular amide--DMSO first peak relative to that of the amide--amide (Table\,\ref{tb:coordCoMamide}) and concomitant slightly enhanced definition of the second and third peaks in the mixtures. In both systems, the first peak of the CoM--CoM RDF, associated to the homomolecular first solvation shell of DMF (Figure\,\ref{fgr:g(r)CoM}, left) and DMAc (Figure\,\ref{fgr:g(r)CoM}, right) shifts to marginally longer distances in the presence of DMSO. However, when interpreting these rather nuanced effects, we must bear in mind that the CoM for each molecule (Figure\,\ref{fgr:dipoles}) is in a different position relative to the C=O group, and that they possess different, albeit comparable, molecular volumes (Table\,\ref{tb:TDparameters}). The $N_{CoM-CoM}(r)$ for the pure liquids show that at the distance of the first minimum of $g_{CoM-CoM}(r)$  DMF and DMAc are surrounded by an average of 13.4 and 13.0 neighbours respectively. When in a 50\,:\,50 mixture the composite coordination shells are made up of $\sim$ 7 amide molecules, and $\sim$ 6.5 DMSO, giving total coordination numbers of 13.7 and 13.9 for DMF and DMAc respectively. Our data therefore indicate that the total coordination shells in the mixtures are remarkably similar to those in the pure amides, and that DMF and DMAc are almost equally solvated by other molecules of the same species as by DMSO. This contrasts with NMF/DMSO mixtures, where there was a strong preference for NMF--DMSO contacts due to N--H\,$\cdots$\,O hydrogen bonding  \cite{cordeiro2011investigation}. 

The CoM--CoM RDFs and cumulative coordination numbers reflect the radially averaged packing structure of the solutions. By interrogating the EPSR model, we are also able to extract the 3--dimensional CoM--CoM spatial density functions (SDFs). These typically reveal whether there is any directional preference within the local coordination environments. With this in mind, Figure\,\ref{fgr:CoM-SDFs} presents the CoM--CoM SDFs for the pure liquids and mixtures. These functions show that the first coordination shell is distributed broadly over the molecular spheroid, particularly over the C=O groups and the O=C--N backbone. As one might expect, lacunae occur over the methyl groups ($\mathrm{H^{A}}$, $\mathrm{H^{Z}}$ and $\mathrm{H^{E}}$). However, we see no clear preference for DMSO over DMF/DMAc, except in the region of the C=O groups. We will examine this effect in detail by analysing the atomic site specific RDFs, coordination numbers and SDFs.

\begin{figure}[h!]
	\includegraphics[width=\columnwidth]{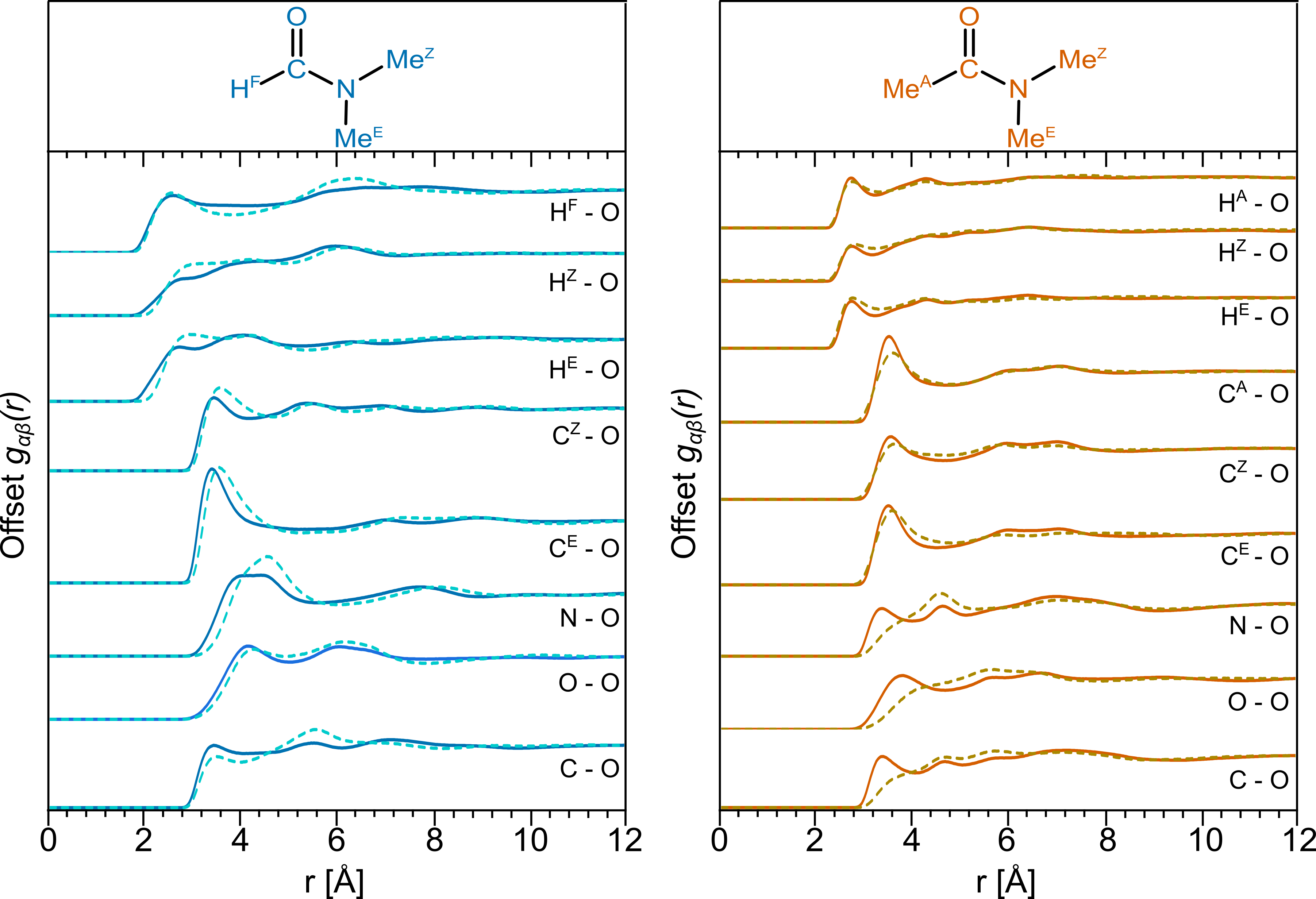}
	\caption{Amide--Amide inter--molecular partial radial distribution functions, $g_{\alpha \beta}(r)$, for DMF (left) and DMAc (right): pure liquids (solid--lines) compared to the structure of DMF and DMAc in the equimolar mixtures with DMSO (dashed--lines). Note the remarkable similarity between the pure and mixture functions for DMF--DMF, and the slight outward shift on mixing for the N--O, O--O and C--O distributions in the DMAc system.}
	\label{fgr:g(r)amide}
\end{figure}
\begin{table}[h]
	
	\centering
	\resizebox{\textwidth}{!} {
		\begin{threeparttable}
			\centering
		\begin{tabular}{lcccc}
			\toprule
			& & \makecell{Peak Position /\AA \\ Pure/Mix}& \makecell{Integration Limit /\AA \\ Pure/Mix} & \makecell{ Coordination Number ($\mathrm{\pm} 0.1$) \\ Pure/Mix}\\
			\midrule
			\multirow{8}*{\color{cerulean}\rotatebox[origin=c]{90}{\textbf{DMF}}}
			& \color{cerulean} ${\mathbf{H^{F}}\textbf{ - O}}$ & 2.53/2.54  & 3.88$^{*}$/3.88$^{*}$	& 1.3/0.6\\
			& \color{cerulean} $\mathbf{H^{Z}}\textbf{ - O}$ &2.70/2.97	&3.53$^{*}$/3.53$^{*}$& 0.7/0.4  \\
			& \color{cerulean} $\mathbf{H^{E}}\textbf{ - O}$ & 2.68/2.86	& 3.63$^{*}$/3.63$^{*}$& 1.1/0.6  \\
			
			& \color{cerulean} $\mathbf{C^{Z}}\textbf{ - O}$ &3.44/3.56	&4.20/4.64& 1.4/1.3 \\
			& \color{cerulean} $\mathbf{C^{E}}\textbf{ - O}$ &3.39/3.54	&5.30/5.17& 4.2/2.1 \\
			
			& \color{cerulean} ${\textbf{N - O}}$ & 4.22/4.56 & 5.70/6.05	& 5.1/3.3\\
			& \color{cerulean} ${\textbf{O - O}}$ & 4.14/4.24 & 5.00/4.93	& 2.8/1.3\\
			& \color{cerulean} ${\textbf{C - O}}$ & 3.40/3.47 & 4.10/3.97 	& 1.3/0.4 \\
			\midrule
			\multirow{9}*{\color{orange}\rotatebox[origin=c]{90}{\textbf{DMAc}}}
			& \color{orange} $\mathbf{H^{A}}\textbf{ - O}$ & 2.71/2.75  & 3.34$^{*}$/3.34$^{*}$	& 1.4/0.8\\
			& \color{orange} $\mathbf{H^{Z}}\textbf{ - O}$ &2.73/2.76	&3.26$^{*}$/3.26$^{*}$& 1.0/0.6  \\
			& \color{orange} $\mathbf{H^{E}}\textbf{ - O}$ & 2.71/2.73	& 3.29$^{*}$/3.29$^{*}$& 1.3/0.9  \\
			
			& \color{orange} $\mathbf{C^{A}}\textbf{ - O}$ &3.50/3.57	&4.74/4.80& 2.1/1.1 \\
			& \color{orange} $\mathbf{C^{Z}}\textbf{ - O}$ &3.53/3.62	&4.64/4.70& 1.7/1.1 \\	
			& \color{orange} $\mathbf{C^{E}}\textbf{ - O}$ &3.50/3.57	&4.45/4.83& 1.6/1.3 \\
			
			& \color{orange} ${\textbf{N - O}}$ & 3.34/3.79 & 4.00/5.31	& 0.8/1.7\\
			& \color{orange} ${\textbf{O - O}}$ & 3.78/4.42 & 4.73/4.86	& 1.8/0.9\\
			& \color{orange} ${\textbf{C - O}}$ & 3.35/3.76 & 4.06/4.17	& 0.9/0.3 \\
			
			\bottomrule

	\end{tabular}
	\caption{Amide--amide peak positions, integration limits and coordination numbers for interactions between the main sites of interest for DMF and DMAc in the pure liquids and the 50\,:\,50 mixture with DMSO.}
	\label{tb:coord-amide}
	        \begin{tablenotes}
		\item[$*$] The integration limit is set to the first minimum of the mix to allow more direct comparison between the corresponding H-bonding coordination numbers.

	\end{tablenotes}
\end{threeparttable}
}
\end{table}

\begin{figure}[h]
	\includegraphics[width=\columnwidth]{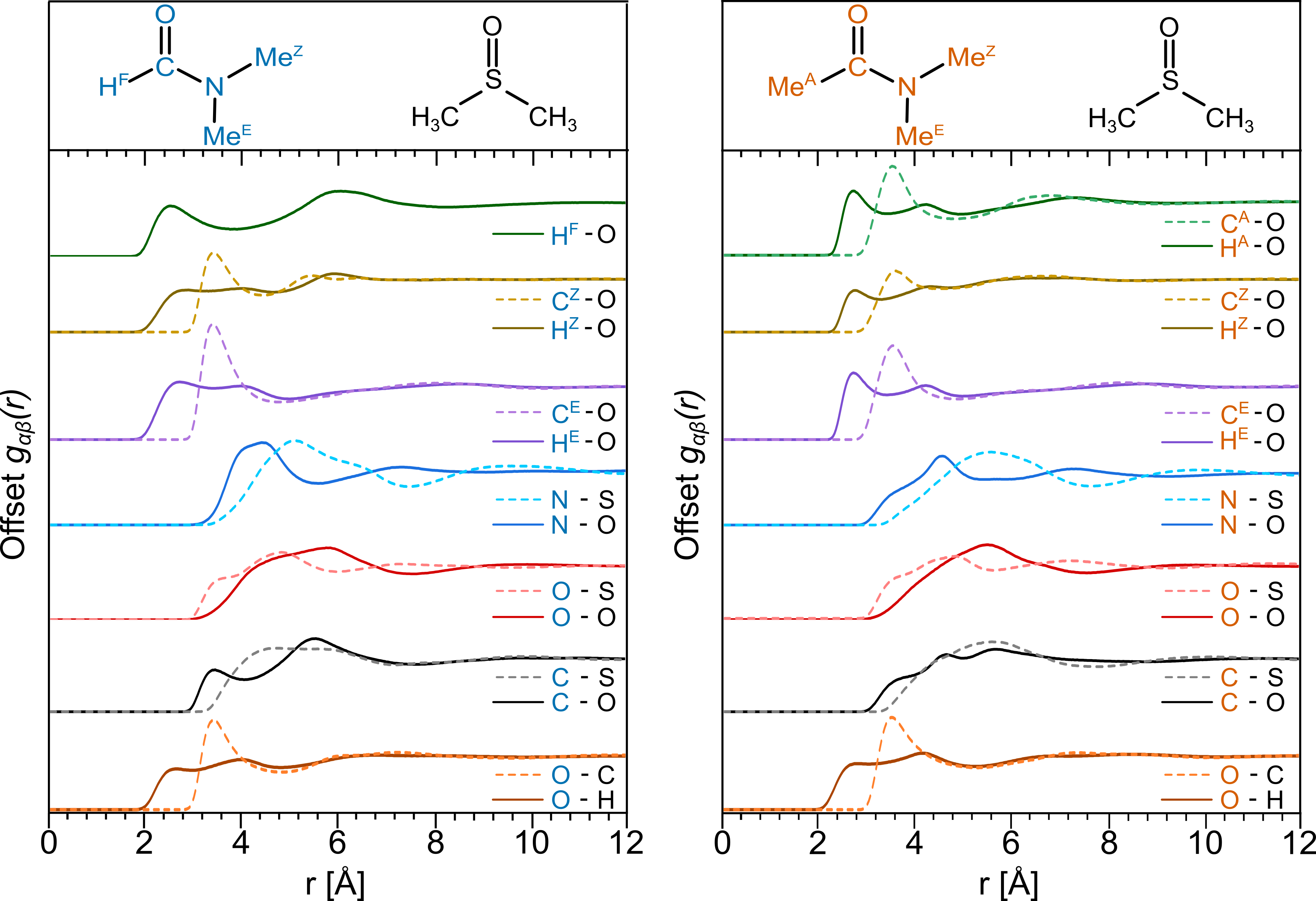}
	\caption{Amide-DMSO inter-molecular partial radial distribution functions, $g_{\alpha \beta}(r)$, for DMF/DMSO (left) and DMAc/DMSO (right). Note that the approaches between DMF and DMAc with DMSO are very similar, the main differences lie in the shorter inter-atomic distances in case of DMF and is linked to the ability of the formic proton $\mathrm{H^{F}}$ to interact via C--H\,$\cdots$\,O hydrogen bonding.}
	\label{fgr:g(r)amide-dmso}
\end{figure}
To understand the inter--molecular amide--amide interactions between two specific atomic sites, we can extract the relevant partial radial distribution $g_{\alpha \beta}(r)$. Figure\,\ref{fgr:g(r)amide} reports selected $g_{\alpha \beta}(r)s$ for DMF (left) and DMAc (right) in both the pure liquid (solid--line) and the mixture with DMSO (dashed--line). The corresponding intermolecular partial structure factors, $S_{\alpha\beta}(Q)$, are presented in Supporting Information Figure S6. Table\,\ref{tb:coord-amide} provides the peak positions and coordination numbers, along with the integration limits. Note that in Table\,\ref{tb:coord-amide}, we have used the same integration limit in the pure and mixed systems for the H--O correlations. Further pairs are given in Supporting Information Figure\,S3\,and\,S4.

First, we focus on the amide--amide interactions between the various protons and the O(=C) site, $g_{H-O}(r)$. In DMF, the $\mathrm{H^{F}}$--to--oxygen distribution shows a peak at $\sim$\,2.5\,\AA, which is consistent with a weak hydrogen bond of electrostatic nature \cite{jeffrey1997introduction}. The coordination numbers for this pair of sites in the pure and mixed systems are 1.3 and 0.6 respectively, in line with the overall mole fraction of DMF and indicative of no preference for/against DMSO. In both pure DMF and DMAc, the methyl protons $\mathrm{H^{A}}$, $\mathrm{H^{Z}}$ and $\mathrm{H^{E}}$ show maxima in the RDF at around 2.7\,\AA. This again is indicative of weak hydrogen bonding. Changes to the RDFs on mixing are subtle, but in the case of DMF we observe peak shifts to $\sim$\,2.9\,\AA\,($\mathrm{H^{Z}}$) and $\sim$\,3.0\,\AA \,($\mathrm{H^{E}}$) in the presence of DMSO.

Turning now to the amide--amide O=C--N backbone correlations, in Figure\,\ref{fgr:g(r)amide} we plot the N--O, O--O and C--O RDFs for both the pure and mixed DMF and DMAc systems. In the case of DMF--DMF, we observe N--O and O--O correlations with peak positions in the range 4.1\,--\,4.6\,\AA, with relatively mild perturbations when comparing the RDFs for the pure and mixed systems. As a general point, however, we note that such discrepancies between specific site-site correlations are not captured by the CoM-CoM RDF shown in Figure\,\ref{fgr:g(r)CoM}. The DMF--DMF $g_{C-O}(r)$s have a first peak at $\sim$ 3.4\,\AA, consistent with the observed H(--C)\,$\cdots$\,O hydrogen bond. From Figure\,\ref{fgr:g(r)amide} and Table\,\ref{tb:coord-amide} we see that nearest neighbour DMAc--DMAc backbone interactions have peak positions in the range 3.3\,--\,3.8\,\AA, slightly shorter than their DMF--DMF counterparts. This confirms the importance of dispersion forces and the steric hindrance of the acetic methyl in pure DMAc.\cite{basma2019liquid} In contrast to the case of DMF, DMAc--DMAc backbone interactions are significantly displaced to longer distances in the presence of DMSO. We can investigate the origins of this effect by turning to the heteromolecular amide--DMSO interactions. 

\begin{table}[h!]
	
	\centering
	\resizebox{\textwidth}{!} {
		\begin{tabular}{lcccc}
			\toprule
			& & Peak Position /\AA& Integration Limit /\AA &Coordination Number ($\mathrm{\pm}\,0.1$)\\
			\midrule
			\multirow{13}*{\color{cerulean}\rotatebox[origin=c]{90}{\textbf{DMF}}}
			
			& \color{cerulean} ${\textbf{H}}\color{black}\textbf{ - O}$ & 2.49  & 3.87	& 0.6\\
			& \color{cerulean} $\mathbf{H^{E}}\color{black}\textbf{ - O}$ &2.64	&3.40& 0.5  \\
			& \color{cerulean} $\mathbf{C^{E}}\color{black}\textbf{ - O}$ &3.40	&4.80& 1.6 \\
			& \color{cerulean} $\mathbf{H^{Z}}\color{black}\textbf{ - O}$ &2.78	&3.28& 0.3  \\
			& \color{cerulean} $\mathbf{C^{Z}}\color{black}\textbf{ - O}$ &3.41	&4.44& 1.0 \\

			& \color{cerulean} ${\textbf{N}}\color{black}\textbf{ - S}$ & 5.12 & 7.45	& 6.6\\
			& \color{cerulean} ${\textbf{N}}\color{black}\textbf{ - O}$ & 4.46 & 5.62	& 2.6\\
			
			& \color{cerulean} ${\textbf{O}}\color{black}\textbf{ - S}$ & 3.58 & 3.85	& 0.3\\
			& \color{cerulean} ${\textbf{O}}\color{black}\textbf{ - O}$ & 5.78 & 7.50	& 6.9\\
			
			& \color{cerulean} ${\textbf{C}}\color{black}\textbf{ - S}$ & 4.56 & 7.18	& 5.9\\
			& \color{cerulean} ${\textbf{C}}\color{black}\textbf{ - O}$ & 3.43 & 4.10 	& 0.5\\
			
			& \color{cerulean} ${\textbf{O}}\color{black}\textbf{ - H}$ & 2.81 & 3.34	& 2.2\\
			& \color{cerulean} ${\textbf{O}}\color{black}\textbf{ - C}$ & 3.51 & 5.20	& 3.9\\

			\midrule
			\multirow{14}{*}{\color{orange}\rotatebox[origin=c]{90}{\textbf{DMAc}}} 	
			
			& \color{orange} ${\mathbf{H^{A}}}\color{black}\textbf{ - O}$ & 2.69  & 3.40 & 0.4\\
			& \color{orange} ${\mathbf{C^{A}}}\color{black}\textbf{ - O}$ & 3.51  & 4.77 & 1.2\\
			& \color{orange} $\mathbf{H^{Z}}\color{black}\textbf{ - O}$ & 2.72 & 3.30 & 0.3  \\
			& \color{orange} $\mathbf{C^{Z}}\color{black}\textbf{ - O}$ &3.56 &4.43 & 0.8 \\		
			& \color{orange} $\mathbf{H^{E}}\color{black}\textbf{ - O}$ &2.70 &3.40 & 0.4  \\
			& \color{orange} $\mathbf{C^{E}}\color{black}\textbf{ - O}$ &3.51 & 4.80 & 1.4 \\

			& \color{orange} ${\textbf{N}}\color{black}\textbf{ - S}$ & 5.56 & 7.65	& 6.4\\
			& \color{orange} ${\textbf{N}}\color{black}\textbf{ - O}$ & 4.57 & 5.28	& 1.7\\
			& \color{orange} ${\textbf{O}}\color{black}\textbf{ - S}$ & 4.83 & 5.55	& 2.0\\
			& \color{orange} ${\textbf{O}}\color{black}\textbf{ - O}$ & 5.50 & 7.41	& 5.9\\

			& \color{orange} ${\textbf{C}}\color{black}\textbf{ - S}$ & 5.60 & 7.76	& 6.7\\
			& \color{orange} ${\textbf{C}}\color{black}\textbf{ - O}$ & 3.70 & 4.04 & 0.3\\

			& \color{orange} ${\textbf{O}}\color{black}\textbf{ - H}$ & 2.79 & 3.32	& 1.9\\
			& \color{orange} ${\textbf{O}}\color{black}\textbf{ - C}$ & 3.43 & 4.80	& 2.6\\
			
			\bottomrule
			
	\end{tabular}}
	
	\caption{Amide--DMSO peak positions, integration limits, and coordination numbers with reference to the main sites of interest on DMF and DMAc and DMSO.}
	\label{tb:coord-dmso}
\end{table}
Selected amide--DMSO $g_{\alpha \beta}(r)$ partial RDFs are shown in Figure\,\ref{fgr:g(r)amide-dmso} for DMF/DMSO (left) and DMAc/DMSO (right) respectively, along with the corresponding peak positions and coordination numbers in Table\,\ref{tb:coord-dmso}. Further pairs are given in supporting information Figure\,S5.  

Regarding the heteromolecular hydrogen bonding, we note that for DMF, the formic proton $\mathrm{H^{F}}$ to DMSO oxygen distribution shows a peak at $\sim$ 2.5\,\AA. This is an indication of a weak, electrostatic hydrogen bond, with C(=O)--H\,$\cdots$\,O=S distance almost identical to that observed for C(=O)--H\,$\cdots$\,O=C between two DMF molecules (Table\,\ref{tb:coord-amide}). Weak hydrogen bonds are also observed between amide methyl protons (H$\mathrm{^{A}}$, H$\mathrm{^{Z}}$ and H$\mathrm{^{E}}$) and DMSO oxygen, and DMSO methyl protons and amide oxygen, at distances between $\sim$\,2.6 and 2.8\,\AA. This methyl C--H\,$\cdots$\,O interaction is extremely weak for $\mathrm{sp^{3}}$ carbon \cite{desiraju2001weak} and in this case is facilitated by the high molecular dipoles of DMF, DMAc and DMSO.
The partial RDFs relative to the amide O=C--N backbone and the O, S, and C sites of DMSO are presented in Figure\,\ref{fgr:g(r)amide-dmso}, with corresponding peak positions and coordination numbers in Table\,\ref{tb:coord-dmso}. In the case of DMF--DMSO, the N--O, O--O, C--O and O--C partial RDFs show only subtle differences when compared with their amide--amide counterparts. This is consistent with our observation that DMF--DMF backbone RDFs are very similar in the bulk and mixed liquids. We attribute this relative insensitivity to the presence of a formic proton, H$\mathrm{^{F}}$, and consequent C(=O)--H\,$\cdots$\,O hydrogen bond as a dominant structural motif in DMF--DMSO mixtures. This C(=O)--H\,$\cdots$\,O interaction  is absent in DMAc, where we observe only very weak Me--H\,$\cdots$\,O hydrogen bonds and dispersion interactions. In this case, DMAc--DMAc backbone interactions were displaced to longer distances in the mixture. We see faint indications in the DMAc--DMSO RDFs that DMSO may compensate for this effect. Specifically, we point to the lower separation shoulders in  $g_{N-O}(r)$ and $g_{C-O}(r)$ occuring at around 3.4\,\AA. To obtain more detailed insight into the solvation shells, we need to look beyond the radially averaged representations of $g_{\alpha\beta}(r)$ and $N_{\alpha\beta}(r)$ and we therefore turn again to the 3--dimensional spatial density functions (SDFs).

\begin{figure}[h!]
	\centering
	\includegraphics[width=\columnwidth]{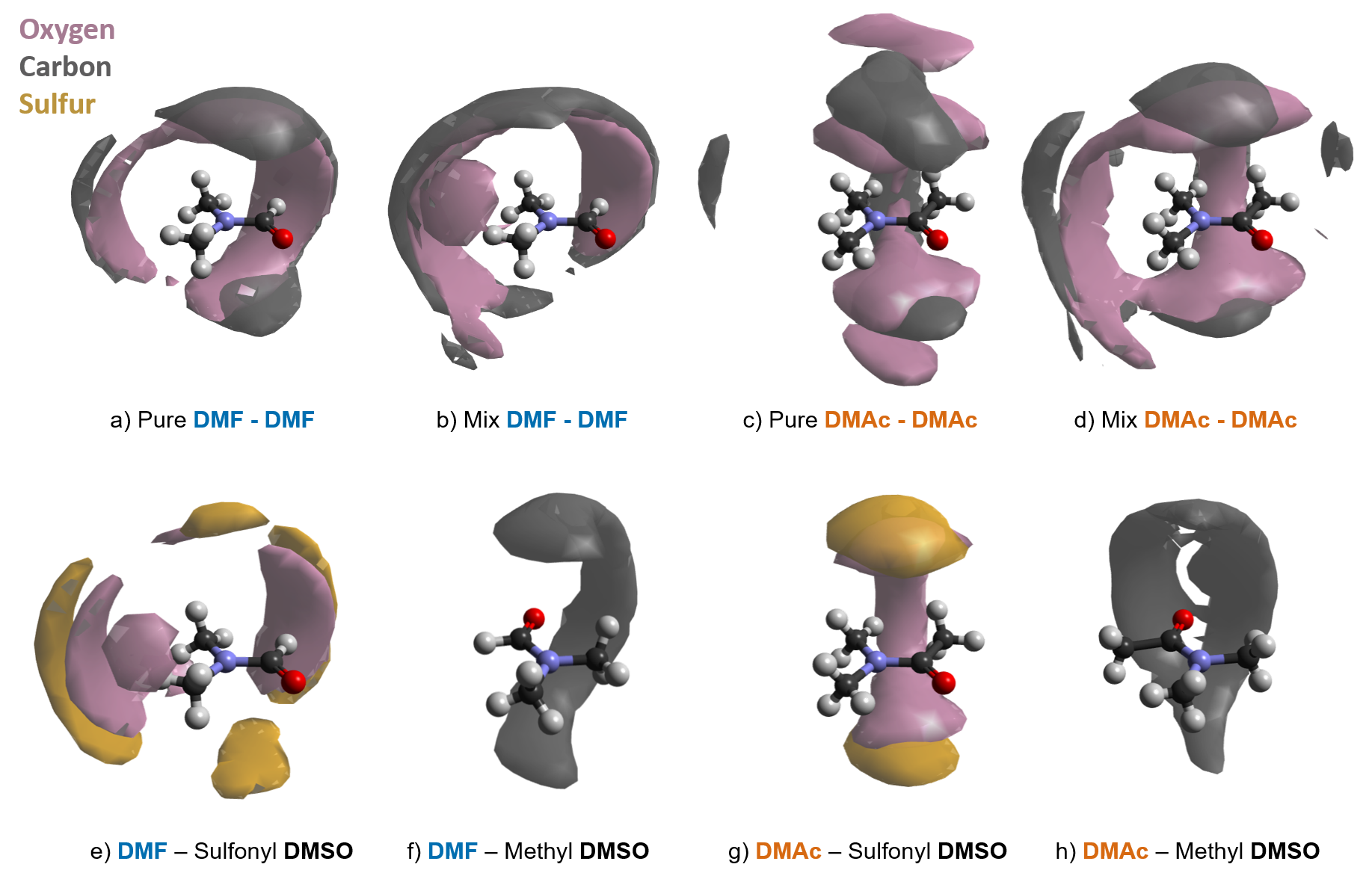}
	\caption{Amide--amide and amide--DMSO spatial density functions (SDFs) representing: (a,b) the 7\% most likely configuration of the DMF C\,(black)\,=\,O\,(pink) group up to 7.5\,\AA\space from the DMF CoM in bulk liquid amide (left) and in the mixture with DMSO (right); (c,d) the 7\% most likely configuration of the DMAc C\,(black)\,=\,O\,(pink) group up to 7.9\,\AA\space from the DMAc CoM in bulk liquid amide (left) and in the mixture with DMSO (right). (e,f) the 5\% most likely position of the DMSO S\,(orange)\,=\,O\,(pink) group up to 7.35\,\AA\space and 7.6\,\AA\space distance from the DMF and DMAc CoM respectively; (g,h) the 5\% most likely position of the DMSO methyl carbons (black) up to 7.35 and 7.6\,\AA\space distance from the DMF and DMAc CoM respectively. Note that Figures (f,h) have been rotated for clarity.}

	\label{fgr:amide-sdf}
\end{figure}

Site--specific spatial density functions are displayed for pure amides and amide--DMSO mixtures in Figure\,\ref{fgr:amide-sdf}. In each case, the reference molecule is shown at the CoM origin, and we focus on the 3--dimensional distribution of neighbouring C=O and S=O groups. In these SDFs, carbonyl carbon density is shown in dark grey, oxygen in pink, and sulfur in orange.

The DMF--DMF functions (Figure\,\ref{fgr:amide-sdf}, panels a\,and\,b) show a band of oxygen density, with more distant carbonyl carbon, centred around the formic proton. This feature extends at longer distances towards the methyl Me$\mathrm{^{Z}}$ groups, and, through a narrow strip, towards Me$\mathrm{^{E}}$ and Me$\mathrm{^{Z}}$. These features confirm the presence of C(=O)--H\,$\cdots$\,O and more distant Me--H\,$\cdots$\,O hydrogen bonds. Overall, the pure and mixed DMF systems have broadly similar amide--amide spatial density, as predicted from our analysis of the RDFs. Relatively delicate changes to the DMF SDFs occur on mixing, and are centred on the lobes corresponding to weaker interactions to the methyl groups, away from the formic proton. The DMF--DMSO SDFs (Figure\,\ref{fgr:amide-sdf}, panels e\,and\,f) support our asseveration that local solvation environment is occupied by DMF and DMSO in equal manner in the mixture. We note that the sulfonyl oxygen density tracks the main features of the amide oxygen, and, likewise, the sulfonyl sulfur and carbonyl carbon. The most likely location for DMSO methyl carbons is a band centred around the amide oxygen, indicative again of weak, directionally rather unconstrained, Me--H\,$\cdots$\,O hydrogen bonds.
	
The DMAc--DMAc functions (Figure\,\ref{fgr:amide-sdf}, panels c and d) show clearly contrasting behaviour to those of DMF. In the pure liquid, the most likely DMAc oxygen and carbonyl carbon density is symmetrically above and below the plane of the O=C--N backbone, with a connecting strip directed towards the methyl protons H$\mathrm{^{A}}$ and H$\mathrm{^{E}}$. This picture is fundamentally different to that for DMF shown in Figure\,\ref{fgr:amide-sdf} panels a and b. Also in contrast to DMF, for DMAc there are clear differences in the amide--amide SDFs for the pure and mixed systems. Specifically, Figure\,\ref{fgr:amide-sdf} panel d shows an additional lobe directed towards the methyl protons H$\mathrm{^{Z}}$ and H$\mathrm{^{E}}$, and concomitant loss of density in the features seen in the pure system, panel c. The reason for this behaviour becomes clear when we examine the DMAc--DMSO SDFs. We see that in the mixture, DMSO sulfonyl oxygen and sulfur mimic the closest approach of DMAc carbonyl oxygen and carbon in the pure liquid. We conclude that DMSO prefers to approach DMAc from above and below the plane of the O=C--N backbone, or axially around Me$\mathrm{^{A}}$ and Me$\mathrm{^{E}}$. In doing this, DMSO molecules displaces some of the DMAc--DMAc interactions observed in the pure liquid, thereby displacing the DMAc--DMAc density in the mixture towards Me$\mathrm{^{Z}}$ and Me$\mathrm{^{E}}$. This is entirely consistent with the shifts observed in the DMAc--DMAc backbone RDFs (Figure 6). As with DMF, the most likely location for DMSO methyl carbons is a band centred around the amide oxygen, but broadened and cleft due to the presence of Me$\mathrm{^{A}}$.

\section{Conclusions}
Neutron diffraction augmented by isotopic substitution of hydrogen (H) for deuterium (D) has been used to study pure DMF and DMAc, and equimolar mixtures of DMF in DMSO and DMAc in DMSO in the liquid state. The atomistic resolution provided by neutron diffraction is critical for understanding weak inter--molecular interactions on a molecular level, as these bonding motifs are elusive and often invisible to many experimental techniques. Empirical Potential Structure Refinement (EPSR) has been used to generate 3--dimensional atomistic models that are consistent with the experimental data. This approach has enabled us to uncover individual site--site interactions, and to conduct detailed comparison between the pure and mixed systems. Our scattering data show that the amides and DMSO are mixed on the nanoscale. Analysis of the EPSR Centre--of--Mass (CoM) correlations shows that in all of our systems the coordination shell contains $\sim$\,13\,--\,14 molecules, and that in the mixed systems there is, on average, near ideal solvation shell sharing between amide and DMSO. Examination of site--specific correlations reveals a rich structural landscape, in which replacement of the formic proton H$\mathrm{^{F}}$ in DMF by$  $ the methyl group Me$\mathrm{^{A}}$ in DMAc leads to fundamentally different solvation environments for these amides in both the pure liquids and mixtures. Weak C--H\,$\cdots$\,O hydrogen bonds are formed by formic (DMF) and methyl (DMF,  DMAc) protons, with distances around 2.5\,\AA\space and 3.0\,\AA. In addition, we observe dispersion interactions above and below the plane of the O=C--N amide $\pi$--backbones, particularly in the DMAc systems in which dispersive forces are expected to be predominant. By comparing pure amides with the liquid mixtures, we show that DMSO has a noteworthy ability to share in hydrogen bonding to both formic and methyl groups, matching the bond distances observed in the pure amides leading to similar solvation motifs and perfect mixing. As a result, DMSO is able to penetrate the amide--amide solvation shells while causing only subtle disruption to the amide--amide interactions. The new knowledge provided by our study is particularly important to tailor electrolytes in confined geometries such as battery electrodes and supercapacitors, since molecular mixing and local interactions are likely to impact on performance.

\begin{acknowledgement}
	The authors acknowledge the UK Science and Technology Facilities Council (STFC) for NIMROD beamtime allocation (10.5286/ISIS.E.RB1910503; ISIS.E.RB1610416; ISIS.E.RB1700030) and the use of SCARF computational facility for the EPSR simulations.
	AJC thanks the Ramsay Memorial Fellowship Trust and the Royal Society University Research Fellowships Scheme for funding.
	Engineering and Physical Sciences Research Council (EPSRC, grant EP/R513143/1) is acknowledged for support of a PhD studentship for CDM. 
\end{acknowledgement}

\begin{suppinfo}
	
	The Supporting Information is available free of charge:
	\begin{itemize}
		\item Total Radial Distribution Functions 
		\item Molecular Bond Lengths and Angles
		\item Additional Partial Radial Distribution Functions 
		\item Intermolecular Partial Structure Factors
		\item Classical Monte Carlo Simulations 
	\end{itemize}
	
\end{suppinfo}


\bibliography{biblio_mixtures5}

\providecommand{\latin}[1]{#1}
\makeatletter
\providecommand{\doi}
  {\begingroup\let\do\@makeother\dospecials
  \catcode`\{=1 \catcode`\}=2 \doi@aux}
\providecommand{\doi@aux}[1]{\endgroup\texttt{#1}}
\makeatother
\providecommand*\mcitethebibliography{\thebibliography}
\csname @ifundefined\endcsname{endmcitethebibliography}
  {\let\endmcitethebibliography\endthebibliography}{}
\begin{mcitethebibliography}{55}
\providecommand*\natexlab[1]{#1}
\providecommand*\mciteSetBstSublistMode[1]{}
\providecommand*\mciteSetBstMaxWidthForm[2]{}
\providecommand*\mciteBstWouldAddEndPuncttrue
  {\def\EndOfBibitem{\unskip.}}
\providecommand*\mciteBstWouldAddEndPunctfalse
  {\let\EndOfBibitem\relax}
\providecommand*\mciteSetBstMidEndSepPunct[3]{}
\providecommand*\mciteSetBstSublistLabelBeginEnd[3]{}
\providecommand*\EndOfBibitem{}
\mciteSetBstSublistMode{f}
\mciteSetBstMaxWidthForm{subitem}{(\alph{mcitesubitemcount})}
\mciteSetBstSublistLabelBeginEnd
  {\mcitemaxwidthsubitemform\space}
  {\relax}
  {\relax}

\bibitem[Clancy \latin{et~al.}(2015)Clancy, Melbourne, and
  Shaffer]{clancy2015one}
Clancy,~A.; Melbourne,~J.; Shaffer,~M. A one-step route to solubilised,
  purified or functionalised single-walled carbon nanotubes. \emph{J. Mat.
  Chem. A} \textbf{2015}, \emph{3}, 16708--16715\relax
\mciteBstWouldAddEndPuncttrue
\mciteSetBstMidEndSepPunct{\mcitedefaultmidpunct}
{\mcitedefaultendpunct}{\mcitedefaultseppunct}\relax
\EndOfBibitem
\bibitem[Chen \latin{et~al.}(2012)Chen, Freunberger, Peng, Bard{\'e}, and
  Bruce]{chen2012li}
Chen,~Y.; Freunberger,~S.~A.; Peng,~Z.; Bard{\'e},~F.; Bruce,~P.~G.
  $\mathrm{Li-O_{2}}$ battery with a dimethylformamide electrolyte. \emph{J.
  Am. Chem. Soc.} \textbf{2012}, \emph{134}, 7952--7957\relax
\mciteBstWouldAddEndPuncttrue
\mciteSetBstMidEndSepPunct{\mcitedefaultmidpunct}
{\mcitedefaultendpunct}{\mcitedefaultseppunct}\relax
\EndOfBibitem
\bibitem[Yu \latin{et~al.}(2020)Yu, Huang, Du, Wang, Wang, Wu, and
  Zhang]{yu2020renaissance}
Yu,~Y.; Huang,~G.; Du,~J.-Y.; Wang,~J.-Z.; Wang,~Y.; Wu,~Z.-J.; Zhang,~X.-B. A
  renaissance of N, N-dimethylacetamide-based electrolytes to promote the
  cycling stability of $\mathrm{Li-O_{2}}$ batteries. \emph{Energy Environ.
  Sci} \textbf{2020}, \emph{13}, 3075--3081\relax
\mciteBstWouldAddEndPuncttrue
\mciteSetBstMidEndSepPunct{\mcitedefaultmidpunct}
{\mcitedefaultendpunct}{\mcitedefaultseppunct}\relax
\EndOfBibitem
\bibitem[Miller and Parker(1961)Miller, and Parker]{1961}
Miller,~J.; Parker,~A.~J. Dipolar Aprotic Solvents in Bimolecular Aromatic
  Nucleophilic Substitution Reactions. \emph{J. Am. Chem. Soc.} \textbf{1961},
  \emph{83}, 117--123\relax
\mciteBstWouldAddEndPuncttrue
\mciteSetBstMidEndSepPunct{\mcitedefaultmidpunct}
{\mcitedefaultendpunct}{\mcitedefaultseppunct}\relax
\EndOfBibitem
\bibitem[Heravi \latin{et~al.}(2018)Heravi, Ghavidel, and
  Mohammadkhani]{heravi2018beyond}
Heravi,~M.~M.; Ghavidel,~M.; Mohammadkhani,~L. Beyond a solvent: triple roles
  of dimethylformamide in organic chemistry. \emph{RSC advances} \textbf{2018},
  \emph{8}, 27832--27862\relax
\mciteBstWouldAddEndPuncttrue
\mciteSetBstMidEndSepPunct{\mcitedefaultmidpunct}
{\mcitedefaultendpunct}{\mcitedefaultseppunct}\relax
\EndOfBibitem
\bibitem[Pastoriza-Santos and Liz-Marz{\'a}n(1999)Pastoriza-Santos, and
  Liz-Marz{\'a}n]{pastoriza1999formation}
Pastoriza-Santos,~I.; Liz-Marz{\'a}n,~L.~M. Formation and stabilization of
  silver nanoparticles through reduction by N, N-dimethylformamide.
  \emph{Langmuir} \textbf{1999}, \emph{15}, 948--951\relax
\mciteBstWouldAddEndPuncttrue
\mciteSetBstMidEndSepPunct{\mcitedefaultmidpunct}
{\mcitedefaultendpunct}{\mcitedefaultseppunct}\relax
\EndOfBibitem
\bibitem[Le~Bras and Muzart(2017)Le~Bras, and Muzart]{le2017n}
Le~Bras,~J.; Muzart,~J. \emph{N, N-Dimethylformamide and N, N-dimethylacetamide
  as carbon, hydrogen, nitrogen and/or oxygen sources}; Wiley-VCH: Weinheim,
  Germany, 2017\relax
\mciteBstWouldAddEndPuncttrue
\mciteSetBstMidEndSepPunct{\mcitedefaultmidpunct}
{\mcitedefaultendpunct}{\mcitedefaultseppunct}\relax
\EndOfBibitem
\bibitem[Ding and Jiao(2012)Ding, and Jiao]{ding2012n}
Ding,~S.; Jiao,~N. N, N-dimethylformamide: A multipurpose building block.
  \emph{Angew. Chem. Int. Ed.} \textbf{2012}, \emph{51}, 9226--9237\relax
\mciteBstWouldAddEndPuncttrue
\mciteSetBstMidEndSepPunct{\mcitedefaultmidpunct}
{\mcitedefaultendpunct}{\mcitedefaultseppunct}\relax
\EndOfBibitem
\bibitem[Clancy \latin{et~al.}(2018)Clancy, Bayazit, Hodge, Skipper, Howard,
  and Shaffer]{clancy2018charged}
Clancy,~A.~J.; Bayazit,~M.~K.; Hodge,~S.~A.; Skipper,~N.~T.; Howard,~C.~A.;
  Shaffer,~M.~S. Charged carbon nanomaterials: redox chemistries of fullerenes,
  carbon nanotubes, and graphenes. \emph{Chem. Rev.} \textbf{2018}, \emph{118},
  7363--7408\relax
\mciteBstWouldAddEndPuncttrue
\mciteSetBstMidEndSepPunct{\mcitedefaultmidpunct}
{\mcitedefaultendpunct}{\mcitedefaultseppunct}\relax
\EndOfBibitem
\bibitem[Chen \latin{et~al.}(2017)Chen, Li, Su, Zhang, Chen, Huang, and
  Yu]{chen2017improving}
Chen,~C.; Li,~L.; Su,~J.; Zhang,~C.; Chen,~X.; Huang,~T.; Yu,~A. Improving rate
  capability and reducing over-potential of lithium-oxygen batteries through
  optimization of Dimethylsulfoxide-N/N-dimethylacetamide mixed electrolyte.
  \emph{Electrochim. Acta} \textbf{2017}, \emph{243}, 357--363\relax
\mciteBstWouldAddEndPuncttrue
\mciteSetBstMidEndSepPunct{\mcitedefaultmidpunct}
{\mcitedefaultendpunct}{\mcitedefaultseppunct}\relax
\EndOfBibitem
\bibitem[Mujika \latin{et~al.}(2006)Mujika, Matxain, Eriksson, and
  Lopez]{mujika2006resonance}
Mujika,~J.~I.; Matxain,~J.~M.; Eriksson,~L.~A.; Lopez,~X. Resonance structures
  of the amide bond: The advantages of planarity. \emph{Chem. Eur. J.}
  \textbf{2006}, \emph{12}, 7215--7224\relax
\mciteBstWouldAddEndPuncttrue
\mciteSetBstMidEndSepPunct{\mcitedefaultmidpunct}
{\mcitedefaultendpunct}{\mcitedefaultseppunct}\relax
\EndOfBibitem
\bibitem[Basma \latin{et~al.}(2019)Basma, Cullen, Clancy, Shaffer, Skipper,
  Headen, and Howard]{basma2019liquid}
Basma,~N.; Cullen,~P.; Clancy,~A.; Shaffer,~M.; Skipper,~N.; Headen,~T.;
  Howard,~C. The liquid structure of the solvents dimethylformamide (DMF) and
  dimethylacetamide (DMA). \emph{Mol. Phys.} \textbf{2019}, \emph{117},
  3353--3363\relax
\mciteBstWouldAddEndPuncttrue
\mciteSetBstMidEndSepPunct{\mcitedefaultmidpunct}
{\mcitedefaultendpunct}{\mcitedefaultseppunct}\relax
\EndOfBibitem
\bibitem[Laurence and Gal(2009)Laurence, and Gal]{laurence2009lewis}
Laurence,~C.; Gal,~J.-F. \emph{Lewis basicity and affinity scales: data and
  measurement}; John Wiley \& Sons, 2009\relax
\mciteBstWouldAddEndPuncttrue
\mciteSetBstMidEndSepPunct{\mcitedefaultmidpunct}
{\mcitedefaultendpunct}{\mcitedefaultseppunct}\relax
\EndOfBibitem
\bibitem[Schultz and Hargittai(1993)Schultz, and
  Hargittai]{schultz1993molecular}
Schultz,~G.; Hargittai,~I. Molecular structure of N, N-dimethylformamide from
  gas-phase electron diffraction. \emph{J. Phys. Chem.} \textbf{1993},
  \emph{97}, 4966--4969\relax
\mciteBstWouldAddEndPuncttrue
\mciteSetBstMidEndSepPunct{\mcitedefaultmidpunct}
{\mcitedefaultendpunct}{\mcitedefaultseppunct}\relax
\EndOfBibitem
\bibitem[Schoester \latin{et~al.}(1995)Schoester, Zeidler, Radnai, and
  Bopp]{schoester1995comparison}
Schoester,~P.~C.; Zeidler,~M.~D.; Radnai,~T.; Bopp,~P.~A. Comparison of the
  structure of liquid amides as determined by diffraction experiments and
  molecular dynamics simulations. \emph{Z. Naturforschung A} \textbf{1995},
  \emph{50}, 38--50\relax
\mciteBstWouldAddEndPuncttrue
\mciteSetBstMidEndSepPunct{\mcitedefaultmidpunct}
{\mcitedefaultendpunct}{\mcitedefaultseppunct}\relax
\EndOfBibitem
\bibitem[Cordeiro and Freitas(1999)Cordeiro, and Freitas]{cordeiro1999study}
Cordeiro,~J. M.~M.; Freitas,~L. C.~G. Study of water and dimethylformamide
  interaction by computer simulation. \emph{Z. Naturforschung A} \textbf{1999},
  \emph{54}, 110--116\relax
\mciteBstWouldAddEndPuncttrue
\mciteSetBstMidEndSepPunct{\mcitedefaultmidpunct}
{\mcitedefaultendpunct}{\mcitedefaultseppunct}\relax
\EndOfBibitem
\bibitem[Macchiagodena \latin{et~al.}(2016)Macchiagodena, Mancini, Pagliai, and
  Barone]{macchiagodena2016accurate}
Macchiagodena,~M.; Mancini,~G.; Pagliai,~M.; Barone,~V. Accurate prediction of
  bulk properties in hydrogen bonded liquids: amides as case studies.
  \emph{Phys. Chem. Chem. Phys.} \textbf{2016}, \emph{18}, 25342--25354\relax
\mciteBstWouldAddEndPuncttrue
\mciteSetBstMidEndSepPunct{\mcitedefaultmidpunct}
{\mcitedefaultendpunct}{\mcitedefaultseppunct}\relax
\EndOfBibitem
\bibitem[Park \latin{et~al.}(2009)Park, Min, Lee, Hong, Kim, and
  Lee]{park2009intermolecular}
Park,~S.-K.; Min,~K.-C.; Lee,~C.-K.; Hong,~S.-K.; Kim,~Y.-S.; Lee,~N.-S.
  Intermolecular hydrogen bonding and vibrational analysis of N,
  N-dimethylformamide hexamer cluster. \emph{Bull. Korean Chem. Soc.}
  \textbf{2009}, \emph{30}, 2595--2602\relax
\mciteBstWouldAddEndPuncttrue
\mciteSetBstMidEndSepPunct{\mcitedefaultmidpunct}
{\mcitedefaultendpunct}{\mcitedefaultseppunct}\relax
\EndOfBibitem
\bibitem[Lei \latin{et~al.}(2003)Lei, Li, Pan, and Han]{lei2003structures}
Lei,~Y.; Li,~H.; Pan,~H.; Han,~S. Structures and hydrogen bonding analysis of
  N, N-dimethylformamide and N, N-dimethylformamide- water mixtures by
  molecular dynamics simulations. \emph{J. Phys. Chem. A} \textbf{2003},
  \emph{107}, 1574--1583\relax
\mciteBstWouldAddEndPuncttrue
\mciteSetBstMidEndSepPunct{\mcitedefaultmidpunct}
{\mcitedefaultendpunct}{\mcitedefaultseppunct}\relax
\EndOfBibitem
\bibitem[Xiang \latin{et~al.}(2017)Xiang, Gao, and
  Wu]{doi:https://doi.org/10.1002/9783527805624.ch7}
Xiang,~J.-C.; Gao,~Q.-H.; Wu,~A.-X. \emph{Solvents as Reagents in Organic
  Synthesis}; John Wiley \& Sons, Ltd, 2017; Chapter 7, pp 315--353\relax
\mciteBstWouldAddEndPuncttrue
\mciteSetBstMidEndSepPunct{\mcitedefaultmidpunct}
{\mcitedefaultendpunct}{\mcitedefaultseppunct}\relax
\EndOfBibitem
\bibitem[Awan \latin{et~al.}(2020)Awan, Buriak, Fleck, Fuller, Goltsev, Kerby,
  Lowdell, Mericka, Petrenko, Petrenko, \latin{et~al.}
  others]{awan2020dimethyl}
Awan,~M.; Buriak,~I.; Fleck,~R.; Fuller,~B.; Goltsev,~A.; Kerby,~J.;
  Lowdell,~M.; Mericka,~P.; Petrenko,~A.; Petrenko,~Y., \latin{et~al.}
  Dimethyl sulfoxide: a central player since the dawn of cryobiology, is
  efficacy balanced by toxicity? \emph{Regen. Med.} \textbf{2020}, \emph{15},
  1463--1491\relax
\mciteBstWouldAddEndPuncttrue
\mciteSetBstMidEndSepPunct{\mcitedefaultmidpunct}
{\mcitedefaultendpunct}{\mcitedefaultseppunct}\relax
\EndOfBibitem
\bibitem[Moffitt(1950)]{moffitt1950nature}
Moffitt,~W. The nature of the sulphur-oxygen bond. \emph{Proc. R. Soc. Lond. A
  Math. Phys. Sci.} \textbf{1950}, \emph{200}, 409--428\relax
\mciteBstWouldAddEndPuncttrue
\mciteSetBstMidEndSepPunct{\mcitedefaultmidpunct}
{\mcitedefaultendpunct}{\mcitedefaultseppunct}\relax
\EndOfBibitem
\bibitem[Cruickshank(1961)]{cruickshank19611077}
Cruickshank,~D. 1077. The role of 3 d-orbitals in $\pi$-bonds between (a)
  silicon, phosphorus, sulphur, or chlorine and (b) oxygen or nitrogen.
  \emph{J. Chem. Soc. (Resumed)} \textbf{1961}, 5486--5504\relax
\mciteBstWouldAddEndPuncttrue
\mciteSetBstMidEndSepPunct{\mcitedefaultmidpunct}
{\mcitedefaultendpunct}{\mcitedefaultseppunct}\relax
\EndOfBibitem
\bibitem[Clark \latin{et~al.}(2008)Clark, Murray, Lane, and
  Politzer]{clark2008dimethyl}
Clark,~T.; Murray,~J.~S.; Lane,~P.; Politzer,~P. Why are dimethyl sulfoxide and
  dimethyl sulfone such good solvents? \emph{J. Mol. Model.} \textbf{2008},
  \emph{14}, 689--697\relax
\mciteBstWouldAddEndPuncttrue
\mciteSetBstMidEndSepPunct{\mcitedefaultmidpunct}
{\mcitedefaultendpunct}{\mcitedefaultseppunct}\relax
\EndOfBibitem
\bibitem[Luzar \latin{et~al.}(1993)Luzar, Soper, and
  Chandler]{luzar1993combined}
Luzar,~A.; Soper,~A.; Chandler,~D. Combined neutron diffraction and computer
  simulation study of liquid dimethyl sulphoxide. \emph{J. Chem. Phys.}
  \textbf{1993}, \emph{99}, 6836--6847\relax
\mciteBstWouldAddEndPuncttrue
\mciteSetBstMidEndSepPunct{\mcitedefaultmidpunct}
{\mcitedefaultendpunct}{\mcitedefaultseppunct}\relax
\EndOfBibitem
\bibitem[McLain \latin{et~al.}(2007)McLain, Soper, and
  Luzar]{mclain2007investigations}
McLain,~S.~E.; Soper,~A.~K.; Luzar,~A. Investigations on the structure of
  dimethyl sulfoxide and acetone in aqueous solution. \emph{J. Chem. Phys.}
  \textbf{2007}, \emph{127}, 174515\relax
\mciteBstWouldAddEndPuncttrue
\mciteSetBstMidEndSepPunct{\mcitedefaultmidpunct}
{\mcitedefaultendpunct}{\mcitedefaultseppunct}\relax
\EndOfBibitem
\bibitem[Soper(2012)]{soper2012computer}
Soper,~A.~K. Computer simulation as a tool for the interpretation of total
  scattering data from glasses and liquids. \emph{Mol. Sim.} \textbf{2012},
  \emph{38}, 1171--1185\relax
\mciteBstWouldAddEndPuncttrue
\mciteSetBstMidEndSepPunct{\mcitedefaultmidpunct}
{\mcitedefaultendpunct}{\mcitedefaultseppunct}\relax
\EndOfBibitem
\bibitem[Megyes \latin{et~al.}(2004)Megyes, Bak{\'o}, Radnai, Gr{\'o}sz,
  Hermansson, Probst, \latin{et~al.} others]{megyes2004x}
Megyes,~T.; Bak{\'o},~I.; Radnai,~T.; Gr{\'o}sz,~T.; Hermansson,~K.;
  Probst,~M., \latin{et~al.}  X-ray and neutron diffraction studies and
  molecular dynamics simulations of liquid DMSO. \emph{Phys. Chem. Chem. Phys.}
  \textbf{2004}, \emph{6}, 2136--2144\relax
\mciteBstWouldAddEndPuncttrue
\mciteSetBstMidEndSepPunct{\mcitedefaultmidpunct}
{\mcitedefaultendpunct}{\mcitedefaultseppunct}\relax
\EndOfBibitem
\bibitem[Cordeiro and Soper(2011)Cordeiro, and
  Soper]{cordeiro2011investigation}
Cordeiro,~J.~M.; Soper,~A.~K. Investigation on the structure of liquid
  N-methylformamide--dimethylsulfoxide mixtures. \emph{Chem. Phys.}
  \textbf{2011}, \emph{381}, 21--28\relax
\mciteBstWouldAddEndPuncttrue
\mciteSetBstMidEndSepPunct{\mcitedefaultmidpunct}
{\mcitedefaultendpunct}{\mcitedefaultseppunct}\relax
\EndOfBibitem
\bibitem[Soper and Phillips(1986)Soper, and Phillips]{soper1986new}
Soper,~A.; Phillips,~M. A new determination of the structure of water at 25 C.
  \emph{Chem. Phys.} \textbf{1986}, \emph{107}, 47--60\relax
\mciteBstWouldAddEndPuncttrue
\mciteSetBstMidEndSepPunct{\mcitedefaultmidpunct}
{\mcitedefaultendpunct}{\mcitedefaultseppunct}\relax
\EndOfBibitem
\bibitem[Soper \latin{et~al.}(1997)Soper, Bruni, and Ricci]{soper1997site}
Soper,~A.; Bruni,~F.; Ricci,~M. Site--site pair correlation functions of water
  from 25 to 400 C: Revised analysis of new and old diffraction data. \emph{J.
  Chem. Phys.} \textbf{1997}, \emph{106}, 247--254\relax
\mciteBstWouldAddEndPuncttrue
\mciteSetBstMidEndSepPunct{\mcitedefaultmidpunct}
{\mcitedefaultendpunct}{\mcitedefaultseppunct}\relax
\EndOfBibitem
\bibitem[McLain \latin{et~al.}(2004)McLain, Benmore, Siewenie, Urquidi, and
  Turner]{mclain2004structure}
McLain,~S.~E.; Benmore,~C.~J.; Siewenie,~J.~E.; Urquidi,~J.; Turner,~J.~F. On
  the structure of liquid hydrogen fluoride. \emph{Angew. Chem.} \textbf{2004},
  \emph{116}, 1986--1989\relax
\mciteBstWouldAddEndPuncttrue
\mciteSetBstMidEndSepPunct{\mcitedefaultmidpunct}
{\mcitedefaultendpunct}{\mcitedefaultseppunct}\relax
\EndOfBibitem
\bibitem[Zhou \latin{et~al.}(2016)Zhou, Deng, Zheng, Xu, Ashraf, and
  Yu]{zhou2016evidences}
Zhou,~Y.; Deng,~G.; Zheng,~Y.-Z.; Xu,~J.; Ashraf,~H.; Yu,~Z.-W. Evidences for
  cooperative resonance-assisted hydrogen bonds in protein secondary structure
  analogs. \emph{Sci. Rep.} \textbf{2016}, \emph{6}, 1--8\relax
\mciteBstWouldAddEndPuncttrue
\mciteSetBstMidEndSepPunct{\mcitedefaultmidpunct}
{\mcitedefaultendpunct}{\mcitedefaultseppunct}\relax
\EndOfBibitem
\bibitem[Chebaane \latin{et~al.}(2012)Chebaane, Hammami, Bahri, and
  Nasr]{chebaane2012intramolecular}
Chebaane,~A.; Hammami,~F.; Bahri,~M.; Nasr,~S. Intramolecular and
  intermolecular interactions in N-methylformamide--water mixture: X-ray
  scattering and DFT calculation study. \emph{J. Mol. Liq.} \textbf{2012},
  \emph{165}, 133--138\relax
\mciteBstWouldAddEndPuncttrue
\mciteSetBstMidEndSepPunct{\mcitedefaultmidpunct}
{\mcitedefaultendpunct}{\mcitedefaultseppunct}\relax
\EndOfBibitem
\bibitem[Borges and Cordeiro(2013)Borges, and Cordeiro]{borges2013hydrogen}
Borges,~A.; Cordeiro,~J.~M. Hydrogen bonding donation of N-methylformamide with
  dimethylsulfoxide and water. \emph{Chem. Phys. Lett.} \textbf{2013},
  \emph{565}, 40--44\relax
\mciteBstWouldAddEndPuncttrue
\mciteSetBstMidEndSepPunct{\mcitedefaultmidpunct}
{\mcitedefaultendpunct}{\mcitedefaultseppunct}\relax
\EndOfBibitem
\bibitem[Ludwig \latin{et~al.}(1995)Ludwig, Weinhold, and
  Farrar]{ludwig1995temperature}
Ludwig,~R.; Weinhold,~F.; Farrar,~T. Temperature dependence of hydrogen bonding
  in neat, liquid formamide. \emph{J. Chem. Phys.} \textbf{1995}, \emph{103},
  3636--3642\relax
\mciteBstWouldAddEndPuncttrue
\mciteSetBstMidEndSepPunct{\mcitedefaultmidpunct}
{\mcitedefaultendpunct}{\mcitedefaultseppunct}\relax
\EndOfBibitem
\bibitem[Soper(1996)]{soper1996empirical}
Soper,~A. Empirical potential Monte Carlo simulation of fluid structure.
  \emph{Chem. Phys.} \textbf{1996}, \emph{202}, 295--306\relax
\mciteBstWouldAddEndPuncttrue
\mciteSetBstMidEndSepPunct{\mcitedefaultmidpunct}
{\mcitedefaultendpunct}{\mcitedefaultseppunct}\relax
\EndOfBibitem
\bibitem[Sears(1992)]{sears1992neutron}
Sears,~V.~F. Neutron scattering lengths and cross sections. \emph{Neutron news}
  \textbf{1992}, \emph{3}, 26--37\relax
\mciteBstWouldAddEndPuncttrue
\mciteSetBstMidEndSepPunct{\mcitedefaultmidpunct}
{\mcitedefaultendpunct}{\mcitedefaultseppunct}\relax
\EndOfBibitem
\bibitem[Squires(2012)]{squires1996introduction}
Squires,~G.~L. \emph{Introduction to the Theory of Thermal Neutron Scattering},
  3rd ed.; Cambridge University Press, 2012\relax
\mciteBstWouldAddEndPuncttrue
\mciteSetBstMidEndSepPunct{\mcitedefaultmidpunct}
{\mcitedefaultendpunct}{\mcitedefaultseppunct}\relax
\EndOfBibitem
\bibitem[Terban and Billinge(2021)Terban, and Billinge]{terban2021structural}
Terban,~M.~W.; Billinge,~S.~J. Structural analysis of molecular materials using
  the pair distribution function. \emph{Chem. Rev.} \textbf{2021}, \relax
\mciteBstWouldAddEndPunctfalse
\mciteSetBstMidEndSepPunct{\mcitedefaultmidpunct}
{}{\mcitedefaultseppunct}\relax
\EndOfBibitem
\bibitem[Steele(1963)]{steele1963statistical}
Steele,~W.~A. Statistical mechanics of nonspherical molecules. \emph{J. Chem.
  Phys.} \textbf{1963}, \emph{39}, 3197--3208\relax
\mciteBstWouldAddEndPuncttrue
\mciteSetBstMidEndSepPunct{\mcitedefaultmidpunct}
{\mcitedefaultendpunct}{\mcitedefaultseppunct}\relax
\EndOfBibitem
\bibitem[Hansen and McDonald(1988)Hansen, and McDonald]{hansen1988theory}
Hansen,~J.-P.; McDonald,~I.~R. Theory of simple liquids. \emph{Phys. Today}
  \textbf{1988}, \emph{41}, 89--90\relax
\mciteBstWouldAddEndPuncttrue
\mciteSetBstMidEndSepPunct{\mcitedefaultmidpunct}
{\mcitedefaultendpunct}{\mcitedefaultseppunct}\relax
\EndOfBibitem
\bibitem[Bowron \latin{et~al.}(2010)Bowron, Soper, Jones, Ansell, Birch,
  Norris, Perrott, Riedel, Rhodes, Wakefield, \latin{et~al.}
  others]{bowron2010nimrod}
Bowron,~D.; Soper,~A.; Jones,~K.; Ansell,~S.; Birch,~S.; Norris,~J.;
  Perrott,~L.; Riedel,~D.; Rhodes,~N.; Wakefield,~S., \latin{et~al.}  NIMROD:
  The Near and InterMediate Range Order Diffractometer of the ISIS second
  target station. \emph{Rev. Sci. Instrum} \textbf{2010}, \emph{81},
  033905\relax
\mciteBstWouldAddEndPuncttrue
\mciteSetBstMidEndSepPunct{\mcitedefaultmidpunct}
{\mcitedefaultendpunct}{\mcitedefaultseppunct}\relax
\EndOfBibitem
\bibitem[Soper(2011)]{soper2011gudrunn}
Soper,~A.~K. \emph{GudrunN and GudrunX: programs for correcting raw neutron and
  X-ray diffraction data to differential scattering cross section}; Science \&
  Technology Facilities Council Swindon, UK, 2011\relax
\mciteBstWouldAddEndPuncttrue
\mciteSetBstMidEndSepPunct{\mcitedefaultmidpunct}
{\mcitedefaultendpunct}{\mcitedefaultseppunct}\relax
\EndOfBibitem
\bibitem[Lorentz(1881)]{lorentz1881ueber}
Lorentz,~H. Ueber die Anwendung des Satzes vom Virial in der kinetischen
  Theorie der Gase. \emph{Annalen der physik} \textbf{1881}, \emph{248},
  127--136\relax
\mciteBstWouldAddEndPuncttrue
\mciteSetBstMidEndSepPunct{\mcitedefaultmidpunct}
{\mcitedefaultendpunct}{\mcitedefaultseppunct}\relax
\EndOfBibitem
\bibitem[Halgren(1996)]{halgren1996merck}
Halgren,~T.~A. Merck molecular force field. I. Basis, form, scope,
  parameterization, and performance of MMFF94. \emph{J. Comput. Chem.}
  \textbf{1996}, \emph{17}, 490--519\relax
\mciteBstWouldAddEndPuncttrue
\mciteSetBstMidEndSepPunct{\mcitedefaultmidpunct}
{\mcitedefaultendpunct}{\mcitedefaultseppunct}\relax
\EndOfBibitem
\bibitem[Hanwell \latin{et~al.}(2012)Hanwell, Curtis, Lonie, Vandermeersch,
  Zurek, and Hutchison]{hanwell2012avogadro}
Hanwell,~M.~D.; Curtis,~D.~E.; Lonie,~D.~C.; Vandermeersch,~T.; Zurek,~E.;
  Hutchison,~G.~R. Avogadro: an advanced semantic chemical editor,
  visualization, and analysis platform. \emph{J. Cheminformatics}
  \textbf{2012}, \emph{4}, 1--17\relax
\mciteBstWouldAddEndPuncttrue
\mciteSetBstMidEndSepPunct{\mcitedefaultmidpunct}
{\mcitedefaultendpunct}{\mcitedefaultseppunct}\relax
\EndOfBibitem
\bibitem[Jorgensen and Swenson(1985)Jorgensen, and
  Swenson]{jorgensen1985optimized}
Jorgensen,~W.~L.; Swenson,~C.~J. Optimized intermolecular potential functions
  for amides and peptides. Structure and properties of liquid amides. \emph{J.
  Am. Chem. Soc.} \textbf{1985}, \emph{107}, 569--578\relax
\mciteBstWouldAddEndPuncttrue
\mciteSetBstMidEndSepPunct{\mcitedefaultmidpunct}
{\mcitedefaultendpunct}{\mcitedefaultseppunct}\relax
\EndOfBibitem
\bibitem[Jorgensen \latin{et~al.}(1996)Jorgensen, Maxwell, and
  Tirado-Rives]{jorgensen1996development}
Jorgensen,~W.~L.; Maxwell,~D.~S.; Tirado-Rives,~J. Development and testing of
  the OPLS all-atom force field on conformational energetics and properties of
  organic liquids. \emph{J. Am. Chem. Soc.} \textbf{1996}, \emph{118},
  11225--11236\relax
\mciteBstWouldAddEndPuncttrue
\mciteSetBstMidEndSepPunct{\mcitedefaultmidpunct}
{\mcitedefaultendpunct}{\mcitedefaultseppunct}\relax
\EndOfBibitem
\bibitem[Zheng and Ornstein(1996)Zheng, and Ornstein]{zheng1996molecular}
Zheng,~Y.-J.; Ornstein,~R.~L. A molecular dynamics and quantum mechanics
  analysis of the effect of DMSO on enzyme structure and dynamics: subtilisin.
  \emph{J. Am. Chem. Soc.} \textbf{1996}, \emph{118}, 4175--4180\relax
\mciteBstWouldAddEndPuncttrue
\mciteSetBstMidEndSepPunct{\mcitedefaultmidpunct}
{\mcitedefaultendpunct}{\mcitedefaultseppunct}\relax
\EndOfBibitem
\bibitem[Yan \latin{et~al.}(2017)Yan, Robertson, Tirado-Rives, and
  Jorgensen]{yan2017improved}
Yan,~X.~C.; Robertson,~M.~J.; Tirado-Rives,~J.; Jorgensen,~W.~L. Improved
  description of sulfur charge anisotropy in OPLS force fields: model
  development and parameterization. \emph{J. Phys. Chem. B} \textbf{2017},
  \emph{121}, 6626--6636\relax
\mciteBstWouldAddEndPuncttrue
\mciteSetBstMidEndSepPunct{\mcitedefaultmidpunct}
{\mcitedefaultendpunct}{\mcitedefaultseppunct}\relax
\EndOfBibitem
\bibitem[Soper(2009)]{soper2009inelasticity}
Soper,~A.~K. Inelasticity corrections for time-of-flight and fixed wavelength
  neutron diffraction experiments. \emph{Mol. Phys.} \textbf{2009}, \emph{107},
  1667--1684\relax
\mciteBstWouldAddEndPuncttrue
\mciteSetBstMidEndSepPunct{\mcitedefaultmidpunct}
{\mcitedefaultendpunct}{\mcitedefaultseppunct}\relax
\EndOfBibitem
\bibitem[Jeffrey and Jeffrey(1997)Jeffrey, and
  Jeffrey]{jeffrey1997introduction}
Jeffrey,~G.~A.; Jeffrey,~G.~A. \emph{An introduction to hydrogen bonding};
  Oxford university press New York, 1997; Vol.~12\relax
\mciteBstWouldAddEndPuncttrue
\mciteSetBstMidEndSepPunct{\mcitedefaultmidpunct}
{\mcitedefaultendpunct}{\mcitedefaultseppunct}\relax
\EndOfBibitem
\bibitem[Desiraju and Steiner(2001)Desiraju, and Steiner]{desiraju2001weak}
Desiraju,~G.~R.; Steiner,~T. \emph{The weak hydrogen bond: in structural
  chemistry and biology}; IUCr, Oxford University Press, 2001; Vol.~9\relax
\mciteBstWouldAddEndPuncttrue
\mciteSetBstMidEndSepPunct{\mcitedefaultmidpunct}
{\mcitedefaultendpunct}{\mcitedefaultseppunct}\relax
\EndOfBibitem
\end{mcitethebibliography}
\newpage
\section{Table of Content (TOC) Image}
\begin{figure}[h!]
	\includegraphics[width=8.25cm,height=4.45cm,keepaspectratio]{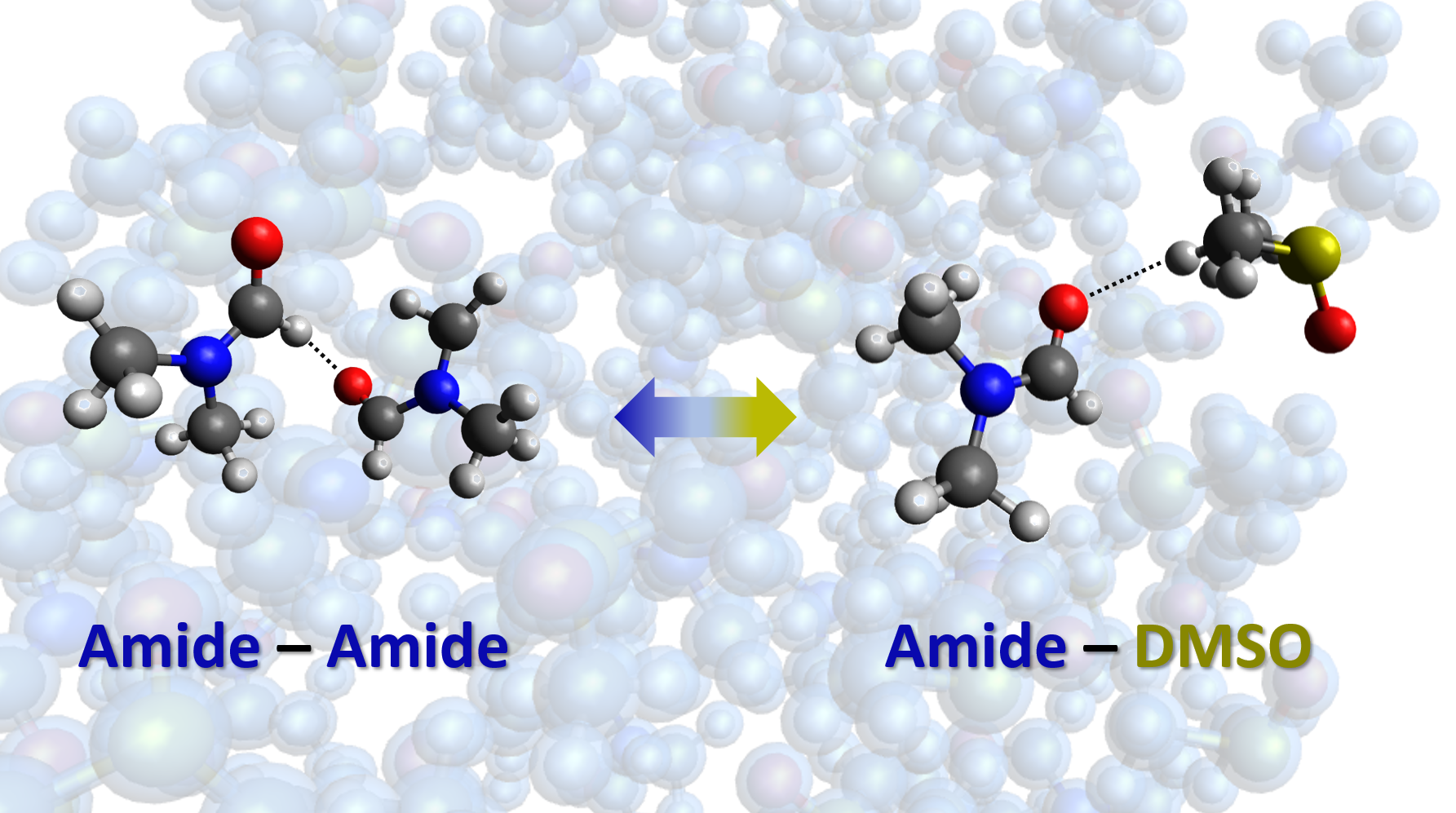}

\end{figure}
\end{document}